\documentclass[twocolumn]{aastex63}

\usepackage[colorinlistoftodos]{todonotes}
\let\Oldtodo\todo
\renewcommand{\todo}[1]{\Oldtodo[inline]{#1}}

\hypersetup{linkcolor=cyan,citecolor=cyan,filecolor=cyan,urlcolor=cyan}

\newcommand{\dblcit}[2]{\cite{#1} and \cite{#2}}

\usepackage{amsmath}
\usepackage{float}

\shorttitle{Sculpting the sub-Saturn Occurrence Rate}
\shortauthors{Hallatt \& Lee}

\graphicspath{{./}{figures/}}

\begin{document}

\title{Sculpting the sub-Saturn Occurrence Rate via Atmospheric Mass Loss}

\correspondingauthor{Tim Hallatt}
\email{thallatt@physics.mcgill.ca}

\author[0000-0003-4992-8427]{Tim Hallatt}
\affil{Department of Physics and McGill Space Institute, McGill University, Montr\'eal, Qu\'ebec, H3A 2T8, Canada}
\affil{Institute for Research on Exoplanets (iREx), Montr\'eal, Qu\'ebec, Canada}

\author[0000-0002-1228-9820]{Eve J.~Lee}
\affil{Department of Physics and McGill Space Institute, McGill University, Montr\'eal, Qu\'ebec, H3A 2T8, Canada}
\affil{Institute for Research on Exoplanets (iREx), Montr\'eal, Qu\'ebec, Canada}

\begin{abstract}
The sub-Saturn ($\sim$4--8$R_{\oplus}$) occurrence rate rises with orbital period out to at least $\sim$300 days. In this work we adopt and test the hypothesis that the decrease in their occurrence towards the star is a result of atmospheric mass loss, which can transform sub-Saturns into sub-Neptunes ($\lesssim$4$R_{\oplus}$) more efficiently at shorter periods. We show that under the mass loss hypothesis, the sub-Saturn occurrence rate can be leveraged to infer their underlying core mass function, and by extension that of gas giants. We determine that lognormal core mass functions peaked near $\sim$10--20$M_{\oplus}$ are compatible with the sub-Saturn period distribution, the distribution of observationally-inferred sub-Saturn cores, and gas accretion theories. Our theory predicts that close-in sub-Saturns should be $\sim$50\% less common and $\sim$30\% more massive around rapidly rotating stars; this should be directly testable for stars younger than $\lesssim$500 Myr. We also predict that the sub-Jovian desert becomes less pronounced and opens up at smaller orbital periods around M stars compared to solar-type stars ($\sim$0.7 days vs.~$\sim$3 days). We demonstrate that exceptionally low-density sub-Saturns, ``Super-Puffs", can survive intense hydrodynamic escape to the present day if they are born with even larger atmospheres than they currently harbor; in this picture, Kepler 223 d began with an envelope $\sim$1.5$\times$ the mass of its core and is currently losing its envelope at a rate $\sim$2$\times 10^{-3}M_{\oplus}~\mathrm{Myr}^{-1}$. If the predictions from our theory are confirmed by observations, the core mass function we predict can also serve to constrain core formation theories of gas-rich planets.
\end{abstract}

\section{Introduction} \label{sec:intro}
 
\begin{figure}
\epsscale{1.2}
\plotone{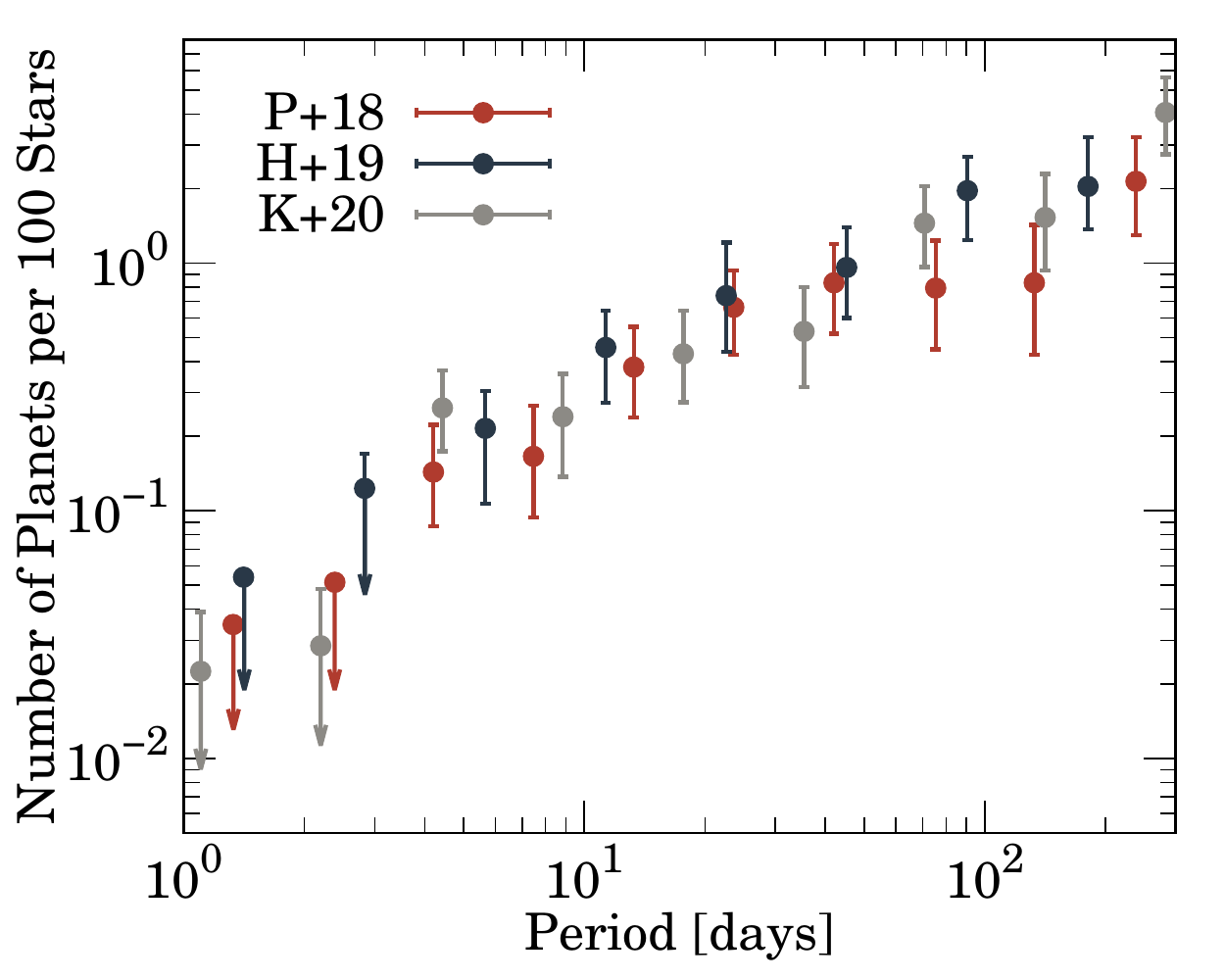}
\caption{The occurrence rate of sub-Saturns, 
as estimated independently by \cite{petmarwin18} (in red, ``P+18"), \cite{hsuforrag19} (in dark navy, ``H+19"), and \cite{kunmat20} (in grey, ``K+20"). \cite{petmarwin18} use stellar parameters and planetary radii from the California Kepler Survey \citep[][]{pethowmar17,johpetful17}, while \cite{hsuforrag19} and \cite{kunmat20} use the Kepler data set augmented with stellar data from Gaia Data Release 2. Error bars indicate $\pm$1$\sigma$ upper and lower limits. Downward arrows indicate 90\% upper limits on the occurrence from \cite{petmarwin18}, and 1$\sigma$ upper limits from \cite{hsuforrag19,kunmat20}. See the text for details about how we plot their data.
\label{figure:subsaturn_occurrence}}
\end{figure}

Massive, gas-rich planets fall into two categories: sub-Saturns ($\sim$4--8$R_{\oplus}$) and Jupiters ($\sim$8--24$R_{\oplus}$). 
Each appear inside orbital distances of $\sim$1 au at a rate of roughly a few percent of solar-type stars \citep[e.g.][]{Howard10,frefrator13,witwanhor20}. 
The occurrence rates of both sub-Saturns and Jupiters
also rise with orbital period out to at at least $\sim$300 days (\citealt{Dong13,petmarwin18,hsuforrag19,kunmat20}; see also Figure \ref{figure:subsaturn_occurrence}),
suggesting that sub-Saturns and Jupiters potentially share a common origin.

In core accretion theory, the cores of gas giants must be sufficiently massive ($\sim$10$M_{\oplus}$) to be capable of triggering ``runaway" gas accretion, rapidly growing into worlds with hundreds of Earth masses of gaseous envelope \citep[e.g.][]{polhubbod96}. Sub-Saturns are failed gas giants; despite having equally massive cores, sub-Saturns have much smaller total masses, having only accreted envelopes tens of percent the mass of their cores \citep[e.g.][]{petsinlop17,milpetbat20}. That sub-Saturns are roughly as common as gas giants is at odds with traditional core accretion theory, which predicted a bimodal planet distribution between gas giants and small, bare cores \citep[e.g.][]{idalin04} (though \cite{lee19} 
recently showed that runaway gas accretion can in fact be readily stymied by disk hydrodynamics). 
This paper does not address the gas accretion histories of sub-Saturns, but rather focuses on the subsequent atmospheric mass loss they suffer after formation ends.
Sub-Saturns' large cross sections and intermediate masses (spanning $\sim$5--50$M_{\oplus}$,) make them susceptible to XUV (X-ray and EUV) photoevaporation and, for sufficiently low-density sub-Saturns, ``boil-off" driven by low gravities and high temperatures \citep{owewu16}. If powerful enough, this can strip Sub-Saturns' envelopes of sufficient mass to transform them into sub-Neptunes ($\lesssim$4$R_{\oplus}$).

The sub-Jovian desert provides strong evidence that sub-Saturns do indeed undergo extra post-processing through atmospheric mass loss compared to their gas giant counterparts. This desert is a roughly triangular region in mass/radius vs.~period space that is virtually devoid of sub-Saturns; demarcated from above by hot Jupiters and from below by hot sub-Neptunes, the desert opens up at $\sim$3 days and widens towards even shorter periods \citep{mazholfai16}.
\cite{matkon16} proposed that the boundaries of the desert reflect tidal circularization of planets on initially highly eccentric orbits (generated by e.g. Kozai \citep[e.g.][]{wumur03}, planet-planet scattering \citep[e.g.][]{rasfor96}, or secular perturbation \citep[e.g.][]{wulit11}), followed by long-term star-planet tidal evolution. In this picture, the desert's triangular morphology is a result of the different mass-radius relations of large and small planets that change how the post-circularization radius scales with planet size. Alternative theories have also been proposed involving in-situ hot Jupiter formation at disk inner edges \citep[for the upper boundary;][]{baibat18} or mass loss of (light) hot Jupiters via photoevaporation and Roche lobe overflow \citep[e.g.][]{kurnak14,valrapras15}. \cite{owelai18} have shown that pure photoevaporative mass loss empties out the desert by stripping sub-Saturns of much of their atmospheres, whose sub-Neptune remnants then comprise the lower boundary of the desert (the upper boundary may be set by tidal circularization of planets undergoing high-eccentricity migration). In stark contrast, Jupiter-mass planets beyond periods $\gtrsim$1 day lose tiny fractions of their atmospheres over Gyr \citep[e.g.][]{murchimur09,kurnak14}.
 
Our investigation does not focus on the sub-Jovian desert; in this paper, we assess the degree to which atmospheric mass loss can sculpt the broader sub-Saturn period distribution out to $\sim$100 days. We adopt and test the hypothesis that the decrease in occurrence towards the star is a result of mass loss of low to intermediate-mass sub-Saturns. We then show that under the mass loss hypothesis, the sub-Saturn occurrence rate can be leveraged to infer their underlying core mass function. 

This paper is organized as follows. In section \ref{sec:methods} we outline our structural and mass loss models, and describe how we construct the sub-Saturn core mass function from their occurrence rate. Section \ref{sec:results} lays out our ``best-fit" core mass distributions, while section \ref{sec:discussion} discusses observational predictions of our theory that can be directly tested in the near future as well as possible limitations of the model. We close with a discussion of broad implications for massive exoplanets.

\section{Methods}\label{sec:methods}

In the theory of photoevaporation, high energy photons from the central star liberate electrons in planetary upper atmospheres which heat the gas to near the escape velocity. This launches an outflowing hydrodynamic wind that sheds the planet's mass and shrinks its radius.
Sub-Saturns that lose enough mass can transform into sub-Neptunes, and this should happen more efficiently closer to the central star. Because the degree of mass loss depends critically on the planet core mass, the sub-Saturn occurrence rate (under the mass loss hypothesis) is thus the fraction of planets with cores massive enough to withstand evaporation and remain sub-Saturns ($\geq$4$R_{\oplus}$). Our goal is to leverage the sub-Saturn occurrence rate to construct their underlying present-day core mass function (integrated over periods $\lesssim$100 days). The decreasing occurrence rate at shorter periods is thus a result of sampling this distribution at larger masses. The advantage of our calculation is that we do not need to perform a population synthesis; in order to produce the sub-Saturn period distribution, we need only be concerned with the required core masses for sub-Saturns to sustain radii $\geq$4$R_{\oplus}$ after mass loss.

To illustrate this, consider an order of magnitude estimate of the mass loss rate in the energy-limited approximation \citep[e.g.][]{watdonwal81,lamselrib03}:

\begin{equation}\label{equation:energy_limited}
    \dot{M}_{\rm env}=-\eta\frac{R^{3}_{\rm p}L_{\rm XUV}}{4a^{2}GM_{\rm p}},
\end{equation}
\noindent
where $\eta\sim$0.15 is the mass loss efficiency (i.e. the fraction of indicent XUV radiation driving mass loss, e.g. \citealt{owejac12,sheionlam14}; see Section \ref{sec:analytic} for more details), $R_{\rm p}$ is the planetary radius, $M_{\rm p}$ is the total planetary mass, $a$ is the orbital semi-major axis, $G$ is the gravitational constant, and $L_{\rm XUV}$ is the stellar high-energy (X-ray plus EUV) luminosity driving the mass loss. Setting the initial planet mass $M_{\rm p}=M_{\rm c}(1+\Gamma)$ with $M_{\rm c}$ the core mass and $\Gamma$ the initial gas to core mass ratio (GCR), the evaporation timescale is $t_{\rm e}\sim M_{\rm c}\Gamma/|\dot{M}_{\rm env}|$. Solving for the core mass yields,

\begin{equation}
M_{\rm c} = \bigg[\frac{\eta R^{3}_{\rm p}L_{\rm XUV} t_{\rm e}}{4a^{2}G\Gamma(1+\Gamma)}\bigg]^{1/2}.
\end{equation}

We define a ``survival" core mass $M_{\rm surv}$ as the core mass below which the planetary envelope evaporates within the $\sim$5 Gyr age of the system (all else being equal). Since $M_{\rm surv}\propto P^{-2/3}R^{3/2}_{\rm p}/\sqrt{\Gamma(1+\Gamma)}$, we find that the survival mass is larger for planets at smaller orbital periods. Photoevaporation theory therefore naturally predicts that the sub-Saturn occurrence rate decreases at small orbital separations. Photoevaporation likely played a similar role in carving out the sub-Jovian desert \citep[][]{owelai18}, setting the maximum radii and masses of remnant sub-Neptunes (i.e., defining the lower edge of the `desert').  Our study on the other hand focuses on the entire sub-Saturn period distribution (out to $\sim$100 days); our goal is to obtain their core mass function defining their occurrence rate profile as the period-dependent survival rate. 

We compute the thermal, mass loss, and structural evolution of our model sub-Saturns with the Modules for Experiments in Stellar Astrophysics (\texttt{MESA}) code, using empirically-derived mass loss and stellar luminosity evolution models. We refine $M_{\rm surv}$ as the survival core mass a sub-Saturn must possess to remain $\geq$4$R_{\oplus}$ after 5 Gyr of mass loss and thermal evolution. The total evolution time is therefore $\sim$5 Gyr. \footnote{While most of the mass loss occurs in the first $\sim$500 Myr, continued cooling contraction of the planet's envelope means that a sub-Saturn at $\sim$500 Myr may not remain a sub-Saturn at $\sim$5 Gyr. The evolution timescale in equation \ref{equation:Msurv} reflects the fact that our calculation of the survival mass also accounts for this thermal contraction. Our mass loss timescale of $\sim$500 Myr is slightly longer than the $\sim$100 Myr quoted by e.g. \cite{owewu13,owewu17}, since the XUV evolution tracks we employ from \cite{johbargud20} yield a slower decrease with time than the power-law models used by \cite{owewu13,owewu17}.} Setting the final planet mass to $M_{\rm c}(1+\Gamma_{\rm F})$, with $\Gamma_{\rm F}$ the final GCR, we may then write,

\begin{equation}\label{equation:Msurv}
    M_{\rm surv} \sim \bigg[\frac{\eta R^{3}_{\rm p}L_{\rm XUV} (5~\mathrm{Gyr})}{4a^{2}G(\Gamma-\Gamma_{\rm F})(1+\Gamma)}\bigg]^{1/2}.
\end{equation}

We measure $M_{\rm surv}$ as a function of planet radius and mass (grouped into ``small" sub-Saturns with initial GCR $\Gamma\sim$0.2, such that after 5 Gyr of thermal evolution without mass loss they end at 5$R_{\oplus}$, and ``large" sub-Saturns with $\Gamma\sim$1, such that they end thermal evolution at 8$R_{\oplus}$), orbital period (from $\sim$1 to $\sim$100 days), and initial stellar rotation period (see Section \ref{sec:stellar_evolution}). We then infer the underlying core mass distribution by calibrating the fraction of cores that lie above $M_{\rm surv}$ at each period against the occurrence rate.

\subsection{Measurements of the sub-Saturn Occurrence Rate}\label{sec:measurements}

We account for three independent estimates of the sub-Saturn occurrence rate: \cite{petmarwin18}, Figure 7; \cite{hsuforrag19}, Table 2, under ``Combined Detection and Vetting Efficiency"; and \cite{kunmat20}, Figure 12. \cite{petmarwin18} use stellar and planet parameters from the California Kepler Survey, which only draws from stars with well characterized spectra \citep[][]{pethowmar17,johpetful17}. Alternatively, \dblcit{hsuforrag19}{kunmat20} study the entire Kepler data set augmented with stellar parameters from Gaia Data Release 2 \citep[][]{GaiaDR2}. \dblcit{hsuforrag19}{kunmat20} use an Approximate Bayesian Computing method to repeatedly simulate Kepler detections of a planet population with a given occurrence rate, and refine this occurrence to minimize a distance function from the observations \citep[see also][for a detailed description]{hsuforrag18}. \cite{petmarwin18} instead maximize a log-likelihood function. In their detection efficiency calculations, \cite{kunmat20} assume all planets have circular orbits, as opposed to \cite{hsuforrag19} who draw eccentricities from a Rayleigh distribution \citep[with a Rayleigh scale parameter of 0.03, as described by][]{hsuforrag18}. 

As shown in Figure \ref{figure:subsaturn_occurrence}, these occurrence estimates generally agree to within $1\sigma$. They differ most between $\sim$40 and 100 days in orbital period, where \dblcit{hsuforrag19}{kunmat20} predict higher median occurrence than \cite{petmarwin18} by roughly a factor of two. In plotting the results of \cite{hsuforrag19}, we sum the median occurrences from 4--6$R_{\oplus}~\mathrm{and} ~ $6--8$R_{\oplus}$ (encompassing the entire sub-Saturn population, defined from 4--8$R_{\oplus}$), adding upper and lower errors in quadrature. Occurrence rate estimates that included a bin with zero or one planet detections are shown as upper limits,  demarcated in Figure \ref{figure:subsaturn_occurrence} with a downward arrow. Occurrence estimates are plotted at the center of bins in logarithm. We multiply the data from \dblcit{hsuforrag19}{kunmat20} by 100 to convert to number of planets per 100 stars.

\subsection{Numerical Method}

We simulate an ensemble of sub-Saturns undergoing thermal evolution and atmospheric mass loss, using the open-source, one-dimensional stellar and planet evolution code $\texttt{MESA}$ \citep[version 11554;][]{paxbildot11,paxcanarr13,paxmarsch15,paxschbau18,paxsmosch19}. In the planetary regime, $\texttt{MESA}$ employs the SCVH equation of state from \cite{saucha95}, and the opacity tables from \cite{fremarlod08}. $\texttt{MESA}$'s built-in planet functionality and microphysics modules are further described in \cite{paxcanarr13}.

Our planets are initialized using the methodology of \cite{cherog16}. They consist of a solid core surrounded by a hydrogen/helium atmosphere. We insert cores less massive than $25M_{\oplus}$ into a $25 M_{\oplus}$ initial gas cloud before scaling the atmosphere to the desired mass via $\texttt{relax\_initial\_mass\_scale}$. For cores $>25M_{\oplus}$, we scale down the 1.05 $M_{\rm Jupiter}$ stock model from the \texttt{irradiated~planet} test suite of \cite{paxcanarr13}. Core radii and bulk densities are computed from the \cite{rogbodlis11} models for rocky composition. Initial entropies are standardized to $s=9~k_{\rm B}~\mathrm{baryon}^{-1}$ using an artificial core luminosity via the \texttt{relax$\_$initial$\_$L$\_$center} routine. 
Initial cooling times (the ratio of the initial internal energy to the luminosity) in the absence of irradiation for typical sub-Saturns in our model, with masses $\sim$15--25$M_{\oplus}$ and periods $\sim$10 days, are $\sim$10 Myr.

Envelopes are heated from both the bottom (from the core) and the top (from the star). We account for the time-dependent core luminosity from thermal inertia (taking the constant volume heat capacity $c_{\rm v}=1~\mathrm{J~K^{-1}~g^{-1}}$ and approximating $\dot{T}_{\rm core}$ as the Lagrangian time derivative of the temperature at the base of the envelope, as implemented by \cite{cherog16}) and radiogenic heating. Unlike \cite{cherog16}, we evolve through time the equilibrium temperature of the planet following stellar bolometric luminosity tracks from \cite{johbargud20}. The outer boundary condition is set using the $F_{\star}-\Sigma_{\star}$ approach outlined in \cite{paxcanarr13}, in which a total flux of $F_{\star}/4$ is distributed in the upper atmosphere through a column density $\Sigma_{\star}$. At every time step we distribute this flux across the uppermost 100 grid cells, where the optical depth to visible light from stellar irradiation reaches unity (i.e., where $\kappa_{\nu}\Sigma_\star \sim$2 assuming the opacity to visible light $\kappa_{\nu} \sim 4 \times 10^{-3}\mathrm{cm^{2}~g^{-1}}$; \citealt{gui10}). \footnote{This outer boundary condition differs from that of \cite{cherog16}, who employ a version of \texttt{MESA}'s stock grey atmosphere $T(\tau)$ relation from \cite{gui10} (modified to apply at fixed outgoing optical depth $\tau$=2/3 instead of a fixed pressure level). Because we are concerned with intense hydrodynamic mass loss, our upper atmospheres must track energy exchange from $PdV$ work, advection from deeper layers of the envelope, and stellar irradiation, so the static $T(\tau)$ solution cannot be applied.} We note that the choice of column depth has a minor impact on our final results. Planetary radii are defined as the thermal photosphere (optical depth $\tau=$2/3 to outgoing thermal radiation), which is typically $\sim$1--2\% larger than the visible photosphere.

\subsection{Stellar XUV Luminosity Evolution}\label{sec:stellar_evolution}

An important addition we make to the methodology of \cite{cherog16} is a treatment of X-ray and EUV luminosity evolution as a function of stellar rotation period (see also \cite{kubvid21}). XUV evolution can vary widely even between stars of similar mass, and is governed mainly by the stellar Rossby number $Ro$, the ratio of the stellar rotation period to the convective turnover time \citep[e.g.][]{tujohgud15,johbargud20}. Young, quickly rotating stars begin with small $Ro$ and maximum X-ray luminosity (which is a roughly constant fraction of the bolometric luminosity during this stage). This corresponds to the ``saturated" phase of XUV evolution, defined when $Ro<Ro_{\rm sat}$, for a saturation threshold Rossby number $Ro_{\rm sat}$ (usually taken to be a fixed value). As stars spin down and $Ro$ grows larger than $Ro_{\rm sat}$, the X-ray luminosity decreases and can be well-described as a power-law via $L_{\rm x}/L_{\rm bol}\propto Ro^{\beta}$, with $L_{\rm bol}$ the bolometric luminosity. Because $Ro_{\rm sat}$ is a fixed value, the transition out of saturation occurs later for lower-mass stars (due to their larger convective envelopes, which lengthen convective turnover times) and for stars with faster initial rotation rates.   
EUV luminosity follows a similar evolution, but falls off with time less rapidly than X rays \citep[e.g.][]{kinwhe20}. 

We use the empirical XUV and bolometric luminosity tracks of \cite{johbargud20}, which are provided as functions of stellar mass and initial rotation period (example X-ray and bolometric luminosity tracks are provided in Figures 11 and 17 respectively of \cite{johbargud20}). \cite{johbargud20} used the stellar evolution models of \cite{spademkim13} to follow the internal structure and compute convective turnover times. Since the sub-Saturn occurrence rate is measured for FGK stars, all results in this work assume a stellar mass $M_{\star}=1M_{\odot}$ unless otherwise stated.

\subsection{Atmospheric Mass Loss}

Sub-Saturns with low planetary gravity and high temperatures (e.g. shortly after disk dispersal when they are hot and inflated) may undergo very rapid escape in the ``boil-off" regime \citep[e.g.][see also \cite{kubvidfos20}, their Figure 1]{owewu16,foserklam17,kubfoserkcub18}. 
Mass loss for more strongly bound planets may be recombination-limited or energy-limited, depending on whether the flow timescale is long compared to the recombination timescale \citep[e.g.][]{owealv16}. To ensure that our conclusions are robust and are not driven by the mass loss model we adopt, we employ two mass loss treatments, described below.

\subsubsection{Analytic Mass Loss Model}\label{sec:analytic}

As a first approximation to the mass loss rate of sub-Saturns, we use the energy-limited approximation, augmented with a simple model of the boil-off process akin to that employed by \cite{owewu16}. Following \cite{cherog16}, the energy-limited photoevaporative mass loss rate is,

\begin{equation}\label{equation:energy_limited}
    \dot{M}_{\rm EL}=-\eta\frac{L_{\rm XUV}R^{2}_{\rm XUV}R_{\rm p}}{4a^{2}GM_{\rm p}K_{\rm tidal}},
\end{equation}
\noindent
where $\eta$ is the height-averaged heating efficiency, $K_{\rm tidal}$ reduces the planet's potential well due to tidal effects from the host star \citep{erkkullam07}, and $R_{\rm XUV}$ is the effective XUV absorption radius, where the atmosphere is optically thick to XUV photons. We take $R_{\rm XUV}\sim R_{\rm p}+H\mathrm{ln}(P_{\rm photo}/P_{\rm XUV})$, with the scale height $H=(k_{\rm B}T_{\rm photo})/(2m_{\rm H}g)$ and $T_{\rm photo},~P_{\rm photo}$ the visible photospheric temperature and pressure. The pressure $P_{\rm XUV}= (m_{\rm H}GM_{\rm p})/(\sigma_{\nu_{0}}R^{2}_{\rm p})$, with $\sigma_{\nu_{0}}=6\times 10^{-18}(h\nu_{0}/13.6~\mathrm{eV})^{-3}\rm ~cm^{2}$ the hydrogen photoionization cross-section for photons at a typical XUV energy $h\nu_{0}=$20 eV \citep{murchimur09}. Typically, $R_{\rm XUV}/R_{\rm p}$$\sim$1.2--1.5. The tidal correction term $K_{\rm tidal}=1-3R_{\rm XUV}/2R_{\rm Hill}+0.5(R_{\rm XUV}/R_{\rm Hill})^{3}$, with $R_{\rm Hill}=a(M_{\rm p}/3M_{\star})^{1/3}$ the Hill radius \citep{erkkullam07}. We assume the efficiency $\eta=0.15$ throughout parameter space. This value is based on \cite{sheionlam14}, who explicitly computed energy deposition rates from XUV radiation and photoelectrons in the thermosphere of a gas giant from first principles. \cite{salschcze16} find similar efficiencies using the \texttt{CLOUDY} microphysics package for planets less massive than $\sim$100$M_{\oplus}$. A mass-dependent efficiency has also been used in previous studies \citep[e.g.][]{owewu17,wu19,rogowe21},
which amounts to order-unity variations. The difference between the energy-limited analytic mass loss model and the empirical mass loss model fitted to hydrodynamic calculations that we describe in the next section are far greater with up to order of magnitude changes to the mass loss rate (see Section \ref{subsec:model_comparison} for more details). Adopting constant efficiency for our analytic mass loss model is therefore sufficient to test the robustness of our results to distinct mass loss treatments.

The lowest density sub-Saturns we consider may undergo brief but intense periods of boil-off at early times. In this case, mass loss is driven by the internal luminosity of the atmosphere. The upper layers of our models are quasi-isothermal (with variations in temperature across the uppermost column density $\Sigma_{\star}\sim 200~\mathrm{g~cm^{-2}}$---roughly the visible photosphere---of $\sim$50--100 K, or $\sim$5--10\% of the equilibrium temperature), so we assume the outflow takes the form of an isothermal Parker wind \citep[][]{par58}:

\begin{equation}
    \dot{M}_{\rm PW}=4\pi R^{2}_{\rm p}\bigg(\frac{P_{\rm surf}}{c_{\rm s}}\bigg)\sqrt{-W_{0}(-f(R_{\rm p}/R_{\rm B}))},
\end{equation}
\noindent
with $P_{\rm surf}$ and $c_{\rm s}$ the pressure and sound speed at the planetary radius (the $\tau=2/3$ surface to the outgoing thermal radiation), $W_{0}(x)$ a real branch of the Lambert W function \citep[][]{valjefcor00,cra04}, $f(x)=x^{-4}\exp(3-4/x)$, and $R_{\rm B}=GM_{\rm p}/2c^{2}_{\rm s}$ the Bondi radius.

We set $\dot{M}=\mathrm{max}(\dot{M}_{\rm EL},\dot{M}_{\rm PW})$, though the Parker wind only ever exceeds the energy-limited rate for the lowest density planets at early times, when atmospheres are hot and inflated. Similar to \cite{owewu16}, the outermost atmospheric layers exhibit negative luminosities during boil-off, indicating that they are expanding, absorbing the internal luminosity, and converting it to $PdV$ work. These layers are much thicker than the stellar heating depth, indicating that the mass loss is driven by internal luminosity. We have verified that the energy advected by the outflow is of order the gravitational binding energy released during this phase of the planet's contraction.

\subsubsection{Empirical Mass Loss Model}\label{sec:empirical}

Next, we consider the mass loss rates from \cite{kubfoserkcub18} (the ``hydro-based approximation", or HBA), which are empirically fit to the grid of 
one-dimensional radiation hydrodynamics models from \cite{kubfoserkjoh18}. These upper atmosphere models extend from the planetary optical photosphere out to the Roche surface. The optical photosphere is assumed to contain only molecular hydrogen at the equilibrium temperature, and differs in radius from the outer boundary of our \texttt{MESA} models at the outgoing thermal photosphere by $1-2\%$. They account for hydrogen dissociation, recombination, and ionization, as well as cooling from Ly$\alpha$ and H$^{+}_{3}$. They assume a constant height-averaged heating efficiency $\epsilon=0.15$ (the fraction of incident stellar XUV converted into thermal energy) and assume upper atmospheres consist entirely of hydrogen. 
The pressure there is computed using a radiative transfer code that accounts for several opacity sources and typically amounts to $\sim$100 mBar, in good agreement with the thermal photospheric pressure in our \texttt{MESA} models. 
The grid is comprised of 7000 models spanning 1 to 39 $M_{\oplus}$ (total mass), 0.002 to 1.3 au, and 1 to 10 $R_{\oplus}$. The mass loss rate is given by,

\begin{equation}
    \dot{M}_{\rm env}=e^{\beta}(F_{\rm XUV})^{\alpha_{1}}\bigg(\frac{a}{\rm au}\bigg)^{\alpha_{2}}\bigg(\frac{R_{\rm p}}{R_{\oplus}}\bigg)^{\alpha_{3}}\Lambda^{k},
\end{equation}
\noindent
with $F_{\rm XUV}$ the XUV flux, and $\beta, \alpha_{1}, \alpha_{2}, \alpha_{3}, k=\xi+\Theta\mathrm{ln}(a/\rm au), \xi, ~\mathrm{and} ~\Theta$ empirically fit to the grid of models. The restricted Jeans parameter $\Lambda$ is \citep[e.g.][]{foserklam17},

\begin{equation}
    \Lambda=\frac{G M_{\rm p}m_{\rm H}}{k_{\rm B}TR_{\rm p}},
\end{equation}
\noindent
with $m_{\rm H}$ the mass of Hydrogen, $k_{\rm B}$ Boltzmann's constant, and $T$ the equilibrium temperature. There are two branches of the HBA, separating planets susceptible to extreme boil-off ($\Lambda$ below the threshold $e^{\gamma}$) and those with more strongly bound atmospheres ($\Lambda\geq e^{\gamma}$). The threshold value is also fit to simulations, and follows

\begin{multline} 
\gamma=[5.564+0.894~\mathrm{ln}(a/\rm au)]^{-1}\times\\
[15.611-0.578~\mathrm{ln}(F_{\rm XUV})+1.537~\mathrm{ln}(a/\rm au)+1.018~\rm ln(R_{\rm p}/R_{\oplus})].
\end{multline}
\noindent
The fitting parameters from \cite{kubfoserkcub18} are summarized in Table \ref{tab:parameters}. The largest planets at very early times in our simulations can lie 1--2$R_{\oplus}$ above the hydrodynamic approximation grid boundary, but the HBA remains accurate in these specific cases (D. Kubyshkina, private communication).
For planets $\geq$39$M_{\oplus}$, we interpolate between models from the same radiation hydrodynamics code used to calibrate the HBA \footnote{The \texttt{MESA} implementation of the hydrodynamic approximation from \cite{kubvidfos20} may be found \dataset[here]{https://zenodo.org/record/4022393}, and the extended grid may be found \dataset[here]{https://zenodo.org/record/4643823}.}. 

The differences between mass loss models are discussed in Section \ref{subsec:model_comparison}, where we also show that the choice of model does not significantly impact our conclusions.

\begin{deluxetable*}{cCCCCCCC}
\tablecaption{HBA Mass Loss Parameters \citep[from~][]{kubfoserkcub18}. \label{tab:parameters}}
\tablecolumns{4}
\tablewidth{0pt}
\tablehead{
\colhead{} &
\colhead{$\beta$} &
\colhead{$\alpha_{1}$} &
\colhead{$\alpha_{2}$} & 
\colhead{$\alpha_{3}$} &
\colhead{$\xi$} &
\colhead{$\Theta$}
}
\startdata
$\Lambda<e^{\gamma}$ & 32.0199 & 0.4222 & -1.7489 & 3.7679 & -6.8618 & 0.0095 \\
$\Lambda\geq e^{\gamma}$ & 16.4084 & 1.0000 & -3.2861 & 2.7500 & -1.2978 & 0.8846 \\
\enddata
\tablerefs{\cite{kubfoserkcub18}}
\end{deluxetable*}

\subsection{Estimating the Core Mass Distribution}\label{subsec:estimation_method}

We define two ensembles of planets that end at 8$R_{\oplus}$ and 5$R_{\oplus}$ after 5 Gyr of thermal evolution in the absence of mass loss (which begin with initial GCRs $\Gamma \sim$1 and $\Gamma\sim$0.2 respectively). These bracket the sub-Saturn population, and we label these ensembles our ``large" (8$R_{\oplus}$) and ``small" (5$R_{\oplus}$) sub-Saturns. Within each ensemble, we search for the survival core mass $M_{\rm surv}$ at which planets remain $\geq$4$R_{\oplus}$ after 5 Gyr of mass loss. Planets with cores above $M_{\rm surv}$ remain in the sub-Saturn regime after mass loss; those with cores below $M_{\rm surv}$ transform into sub-Neptunes and are deleted from the ensemble. We track $M_{\rm surv}$ as a function of orbital period (sampled at nine orbital periods from 1.3--133 days, matching the bins from \cite{petmarwin18}), the ensemble's final radius without mass loss (either 8 or 5$R_{\oplus}$), and initial stellar rotation period (interpolated between $P_{\star}=$1, 5, and 10 days). Planetary mass loss is computed from 10 Myr to 5 Gyr unless otherwise noted.

To estimate $M_{\rm surv}$, we begin by constructing our ensembles of ``large" and ``small" sub-Saturns as defined above. We take an iterative approach to efficiently sweep through parameter space. We first evolve without mass loss 200 \texttt{MESA} models with core masses sampled at every 2$M_{\oplus}$ in the range $[$5,45$]M_{\oplus}$, with GCRs sampled at every 0.02 in the range $[$0.85,1.05$]$ for our large planet ensemble, and in the range $[$0.1,0.3$]$ for our small planet ensemble; these ranges for our initial grid of GCRs were known a priori to yield planetary radii $\sim$8 and 5$R_{\oplus}$. For each core mass, we record the GCR that yields a final planet radius closest to 8 and 5$R_{\oplus}$ after 5 Gyr for large and small sub-Saturns respectively, and add this planet (core mass and GCR) to our ensemble. This GCR differs slightly (at the 2--4\% level) between the different core masses. The uncertainty introduced by our finite grid resolution, $\pm$0.05$R_{\oplus}$, is significantly smaller than the typical Kepler radius measurement error of $\sim$5\%.

The second step of our estimation procedure is to search within those ensembles for the survival core mass. To do so, we re-run each planet in the ensemble now under the combined effect of thermal evolution and mass loss for 5 Gyr. 
The core mass of the planet that ends closest to but larger than 4$R_{\oplus}$ is deemed the ``survival" core mass. Having identified an approximate survival core mass in these first two steps, we repeat steps one and two, but sampling the core masses at every 0.5$M_{\oplus}$ in the range $[M_{\rm surv}$-1$M_{\oplus}$,$M_{\rm surv}$+1$M_{\oplus}]$. Due to our finite grid, the survival planet ends within 4$\pm$0.05$R_{\oplus}$. This procedure is repeated for different stellar masses, and the mass loss steps are repeated using the different mass loss models (analytic vs. empirical) and initial stellar rotation periods. The procedure is also repeated at each period in our grid. This is because the required GCR for a planet to end at 8 or 5$R_{\oplus}$ decreases at smaller periods, since the extra stellar heating increases the radii at a given GCR. In this way, we ensure that our ensembles of planets are indeed in the sub-Saturn regime (4--8$R_{\oplus}$) at each period. Overall, our results for $M_{\rm surv}$ are based on $\sim$2000 simulations in total.

Our goal in this study is to isolate the effect of mass loss on the sub-Saturn occurrence rate. We therefore assume that prior to evaporation, the period distribution of sub-Saturns is flat in a logarithmic sense. Sub-Saturns undergoing disk migration are large enough to carve out deep gaps that will halt their orbital migration as they move inwards, which can naturally produce a logarithmically flat period distribution \citep{hallee20}. Regardless of whether sub-Saturns arrive at their final periods via migration or in-situ formation, we would expect to find fewer planets inside where disks are truncated. We therefore also assume that disk inner edges are truncated at corotation with their host stars. Associating disk inner edges with the corotation radius of young stars can explain the fall-off of the number of sub-Neptunes inside $\sim$10 days \citep[][]{mulpasapa15,leechi17}, and as we will show, also contributes to the fall-off of the number of sub-Saturns. Stellar rotation periods $P_{\star}$ are sampled via an inverse CDF method from the distribution of pre-main sequence rotation periods from NGC 2362 \citep[][filtered to 0.5--1.4$M_{\odot}$; see Figure 2 of \cite{leechi17} for the rotation period distribution]{irwhodaig08}. The pre-mass loss period distribution is determined by sampling 10000 periods randomly in $\log P$ with the minimum period set by $P_{\star}$ and the maximum period at the outer edge of the occurrence bin that we normalize to --- 100 days for \cite{petmarwin18}, 128 for \cite{hsuforrag19}.

The occurrence at a given period is computed by multiplying the probability that sub-Saturns are initially found at that period by the probability that sub-Saturns survive mass loss (i.e. have cores $\geq M_{\rm surv}$). The survival probability at each period is computed as the fraction of cores that are above $M_{\rm surv}$ from a lognormal core mass distribution. The occurrence is computed in logarithmic period intervals, with $M_{\rm surv}$ in each bin the mean of the distribution of $M_{\rm surv}$ from sampled periods (and associated $P_{\star}$) that lie inside each bin. We choose a lognormal core mass function because sub-Saturns have likely undergone (or were on the verge of undergoing) some degree of runaway accretion to accumulate their massive envelopes, and this can only be achieved if they harbor massive enough cores $\sim$10$M_{\oplus}$ for disks with ISM-like opacities \citep[e.g.][]{polhubbod96,raf06,pisyou14}. We therefore expect their underlying core mass function to be peaked at a characteristic, relatively large ($\sim$10$M_{\oplus}$) core mass. Top-heavy distributions (e.g. a positive power-law in core mass) would also satisfy these constraints, but it is unlikely that nature prefers to create more massive cores than small. Rayleigh functions were also tested, but were found to not be compatible with the data. 
Our lognormal distributions take the form,

\begin{equation}\label{equation:lognormal}
 \frac{dN}{d\ln M}\propto\frac{1}{\sigma_{\rm M} \sqrt{2\pi}}\exp\bigg[-\frac{(\ln M - \ln \mu)^{2}}{2\sigma_{\rm M}^{2}}\bigg].   
\end{equation}
\noindent
with $\sigma_{\rm M}$ the standard deviation and $\mu$ the median. Our mass functions run from 1--100$M_{\oplus}$. The occurrence rate is normalized to match the observations at orbital periods of $\sim$100 days. We search by eye for the core mass distributions which best reproduce the general shape of the measured occurrence rate.

\section{Results}\label{sec:results}
\subsection{Analytic vs. Empirical Mass Loss Models}\label{subsec:model_comparison}

\begin{figure}
\epsscale{1.2}
\plotone{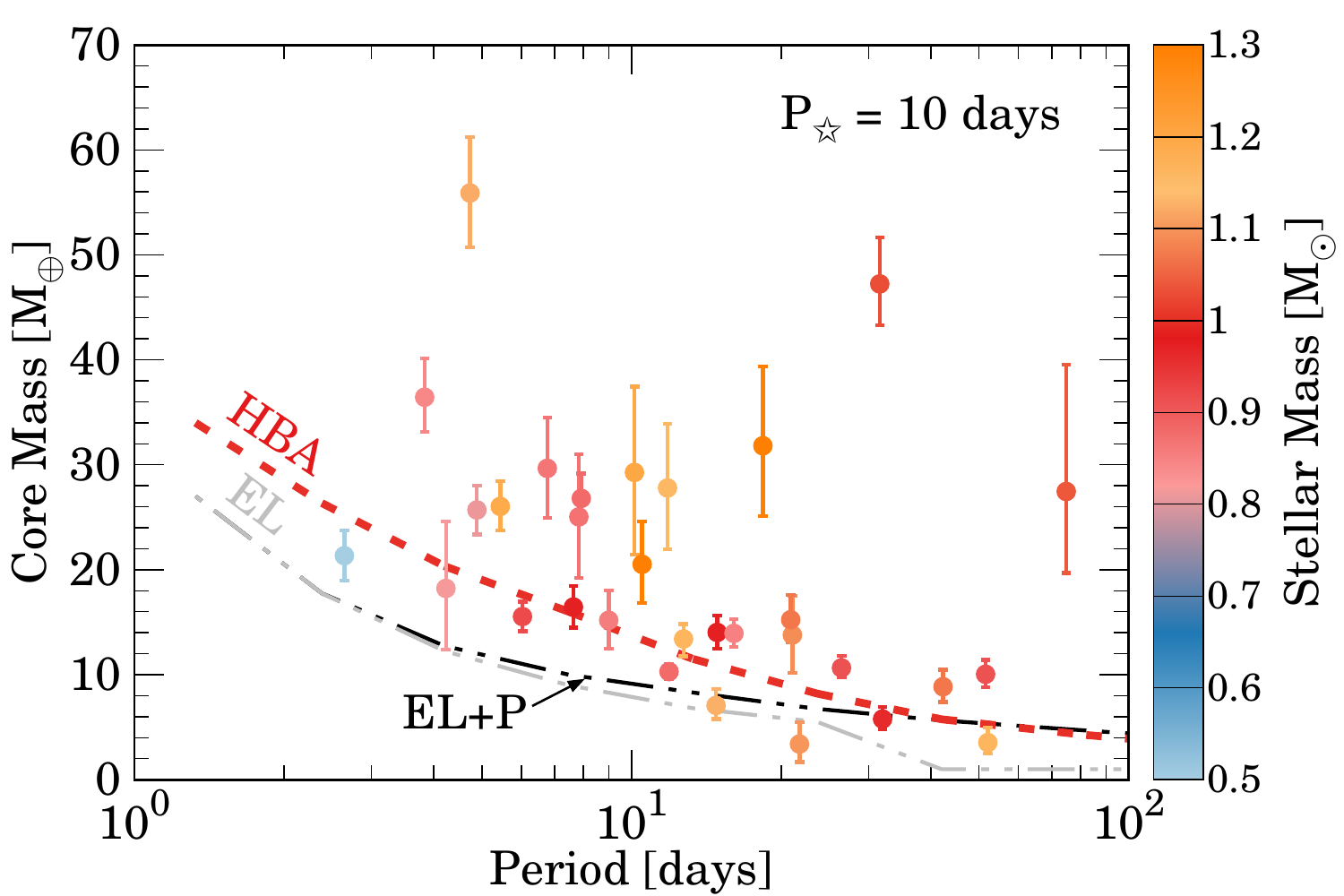}
\caption{
The sub-Saturn survival core mass $M_{\rm surv}$ for large sub-Saturns (i.e. planets with initial gas to core mass ratios $\sim$1, that end at $8R_{\oplus}$ after 5 Gyr of thermal evolution in the absence of mass loss) as a function of orbital period and host stellar mass, computed using different mass loss models: the hydrodynamic approximation (HBA; red dash), energy-limited (EL; gray dash-dot), and energy-limited plus isothermal Parker wind (EL+P; black dash-dot).
Planets in the ensemble with cores lighter than $M_{\rm surv}$ evaporate below 4$R_{\oplus}$, transforming into sub-Neptunes. 
For our model curves, we assume a solar mass star with initial rotation period of 10 days. The EL+P model agrees
with HBA at $\gtrsim$40 days, where mass loss is due mostly to boil-off. The models differ for closer-in planets that are moderately bound ($\Lambda\sim$15--30) (see discussion in Section \ref{subsec:model_comparison}). Estimated core masses of observed sub-Saturns from \cite{milpetbat20} are also shown with their errors. These core masses account for tidal dissipation and are color-coded according to host star mass, which are taken from the \href{https://exoplanetarchive.ipac.caltech.edu}{NASA Exoplanet Archive}. Our theoretical survival core masses for sub-Saturns orbiting solar mass stars trace out the minimum core masses of observed sub-Saturns.
\label{figure:mcritical_HBA_ELP}}
\end{figure}

In order to demonstrate that our model is robust and our choice of mass loss prescription does not drive our results, we begin by comparing the survival core masses computed using the analytic and empirical mass loss models. The survival core mass for our ensemble of large sub-Saturns (i.e. with initial GCRs $\sim$1, such that they end at 8$R_{\oplus}$ after 5 Gyr of thermal evolution) is shown in Figure \ref{figure:mcritical_HBA_ELP}. Planets with core masses below $M_{\rm surv}$ fall below $4R_{\oplus}$ after mass loss, losing $\gtrsim$80\% of their envelopes. We use a stellar mass of 1 $M_{\odot}$ with an initial rotation period $P_{\star}=10$ days. Survival core masses computed using the energy-limited plus Parker wind mass loss model are in black (``EL+P"), results using only the energy-limited approximation are in grey (``EL"), and results using the hydrodynamic approximation are in red (``HBA"). Also shown are estimated core masses for a sample of observed sub-Saturns from \cite{milpetbat20}, color-coded according to host star mass (a comparison between core masses inferred from observations and our model is made in Section \ref{subsec:mass_functions}). 

As illustrated in Figure \ref{figure:mcritical_HBA_ELP}, the analytic (EL+P) and pure energy-limited (EL) models predict similar survival core masses inside $\sim$20 days, but diverge at further distances. Beyond $\sim$20 days, mass loss is dominated by boil-off at early times, which the pure energy-limited model does not account for. Similarly, Figure \ref{figure:mcritical_HBA_ELP} shows that the analytic (EL+P) approach predicts similar survival masses to the empirical (HBA) model at $\gtrsim$40 days where the mass loss is due mostly to boil-off at early times, but predicts lower survival masses at smaller periods. This is because the hydrodynamic approximation and the analytic model differ for planets that are moderately bound. \cite{owewu16} note that the isothermal Parker solution effectively shuts off the wind when $R_{\rm p}<0.1R_{\rm B}$, which is roughly equivalent to planets having $\Lambda>20$ \citep[for isothermal gas; e.g.][]{cuberkjuv17,foserklam17}. The radiation hydrodynamics calculations of \cite{kubfoserkjoh18} resolved the mass loss process without approximating the wind as an isothermal Parker solution, finding that the transition region between boil-off---driven by the internal luminosity and low planetary gravity---and XUV-driven evaporation occurs for 15$\lesssim\Lambda\lesssim$30. Planets in our ensemble inside $\sim$20 days in Figure \ref{figure:mcritical_HBA_ELP} begin at early times in the range $\Lambda\sim$15--30, in the regime where our Parker solution drops off ($R_{\rm p}\sim 0.09 R_{\rm B}$) but the hydrodynamic approximation still indicates that boil-off should dominate over ordinary photoevaporation. Since most of the mass loss occurs in the first tens of Myr of evolution, this sets the difference in final radii.

Mass loss is computed using the empirical (HBA) model for our fiducial results. This choice is motivated by the limitations of our isothermal Parker wind solution, which is imposed without resolving the upper atmosphere (from the photosphere to the Roche surface) as it launches a wind. Our energy-limited treatment of XUV evaporation also neglects the energy exchange through recombination, ionization, dissociation, Ly$\alpha$ and H$_{3}^{+}$ which are accounted for in the hydrodynamic approximation. We stress that our conclusions are only weakly sensitive to the mass loss model we adopt however. As analytic models generally predict smaller survival core masses, the underlying core mass functions that match the observed sub-Saturn occurrence rate are skewed towards lower masses. Fitting the analytic model to \cite{hsuforrag19} requires $\mu=6-7.5M_{\oplus}$ (with $\sigma_{\rm M}=0.7$), as opposed to our fiducial $\mu=9-10M_{\oplus}$ ($\sigma_{\rm M}=0.8$), and fitting the analytic model to \cite{petmarwin18} requires $\mu=10-13M_{\oplus}$ (with $\sigma_{\rm M}=0.7$) instead of our fiducial $\mu=16-21M_{\oplus}$ ($\sigma_{\rm M}=0.7$). The uncertainties associated with the mass loss model, $\pm$3--8$M_{\oplus}$ in $\mu$, are subsumed by the observational uncertainties associated with the occurrence rate (see Section \ref{subsec:mass_functions} for more details).

\begin{figure*}
\centering
\includegraphics[width=1\textwidth]{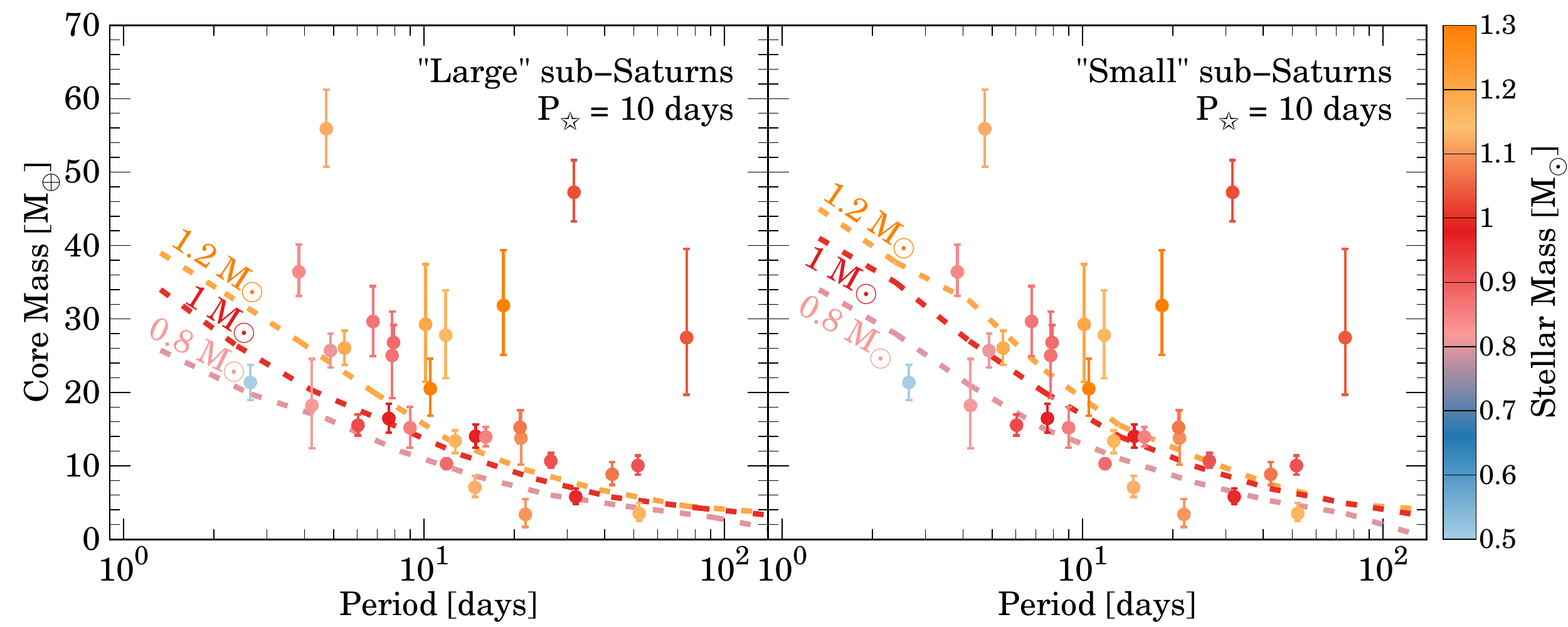}
\caption{
Survival core masses $M_{\rm surv}$ as functions of orbital period and stellar mass, for our ensembles of large and small sub-Saturns (that begin with initial gas to core mass ratios $\sim$1 and $\sim$0.2, in left and right panels, respectively). Planets in each ensemble with cores less massive than $M_{\rm surv}$ lose enough mass to fall below $4R_{\oplus}$, transforming into sub-Neptunes. We show results assuming 0.8, 1, and 1.2$M_{\odot}$ host stars (in pink, red, and orange, respectively) with initial stellar rotation periods of 10 days. Core masses of observed sub-Saturns from \cite{milpetbat20} are plotted and color-coded according to host star mass (stellar masses are from the \href{https://exoplanetarchive.ipac.caltech.edu}{NASA Exoplanet Archive}). The majority of observed sub-Saturns have cores above the survival core mass, in alignment with our model expectations. The two planets inconsistent with the model are Kepler-223 d (at $\sim$15 days) and Kepler-33 d (at $\sim$22 days), both in the ``Super-Puff" regime. The agreement becomes worse for small sub-Saturns (right panel) which are more susceptible to envelope mass loss below 4$R_\oplus$.
\label{figure:mcritical}}
\end{figure*}

\subsection{``Best-fit" Core Mass Functions}\label{subsec:mass_functions}

The survival core masses for our ensembles of large and small sub-Saturns are compared to core masses inferred from observed sub-Saturns in Figure \ref{figure:mcritical} (large sub-Saturns in the left panel, small on the right). We show our model results using stellar masses of 0.8, 1, and 1.2$M_{\odot}$, assuming an initial rotation period $P_{\star}=10$ days. The core masses of observed sub-Saturns from \cite{milpetbat20} are color-coded according to host star mass. \cite{milpetbat20} use the \texttt{MESA} package from \cite{cherog16}, with the addition of a tidal heat source (from eccentricity and obliquity tides, deposited below the radiative-convective boundary) to infer these sub-Saturns' present-day envelope and core masses. The extra heat inflates the planetary radii such 
that some low-density sub-Saturns can be reconciled with smaller envelope masses and therefore more massive cores.
For a given host star mass and assumed initial GCR, cores must lie above the corresponding model curve to remain a sub-Saturn.

Our model for large sub-Saturns successfully recovers all planets except two ``super-puffs" (Kepler 33-d, and Kepler 223-d). We discuss these outliers in Section \ref{sec:super-puffs}. The survival core mass is higher for our small sub-Saturns (that begin with initial GCRs $\sim$0.2), as they only need to lose $\sim$50\% of their atmospheres to fall below 4$R_{\oplus}$. Because of this, our small sub-Saturn curves are inconsistent with four additional planets: HD 219666 b (at $\sim$6 days, a $15.5M_{\oplus}$ core around a $0.92M_{\odot}$ star \citealt{esparmgan19}), Kepler 18 c ($\sim$7.6 days, $16.5M_{\oplus}$ with a $0.97 M_{\odot}$ host \citealt{cocfabtor11}), K2-19 c ($\sim$12 days, $10.3M_{\oplus}$ around a $0.9 M_{\odot}$ star \citealt{mayvanlat18}), and Kepler 25 c ($\sim$13 days, $13.3M_{\oplus}$ around a $1.1 M_{\odot}$ host \citealt{milhowwei19}). Our model therefore predicts that these planets must have begun with initial GCRs $\gtrsim$0.2. 

Kepler 79 d (at 52 days, plotted with a 1.16$M_{\odot}$ host star \citep[][]{jonlisrow14} and a $3.5M_{\oplus}$ core) and Kepler 11 e (at 32 days, around a $0.96M_{\odot}$ host \citep{lisjonrow13} with a $5.8M_{\oplus}$ core) also fall below the theory curve, but nevertheless may not conflict with our model. Using new TTV statistical methodology, \cite{yofofiaha21} recently estimated the mass of Kepler-79 d to be $11.3^{+2.2}_{-2.2}M_{\oplus}$, significantly more massive than the $6.0^{+2.1}_{-1.6}M_{\oplus}$ estimate from \cite{jonlisrow14} used by \cite{milpetbat20} to infer its core mass. This larger total mass implies a larger core mass, large enough to strip Kepler-79d of its super-puff status, which will nudge Kepler-79d's core mass safely above our theoretical survival line (we defer a detailed statistical inference of the new core mass to future work). Similarly, \cite{bedbeamel17} use new stellar measurements to infer a total mass for Kepler 11 e of $9.48^{+0.86}_{-0.88}M_{\oplus}$, larger than the $6.7^{+1.2}_{-1.0}$ from \cite{hadlit17} used by \cite{milpetbat20} to infer its core mass. This larger mass implies a core mass of $\sim$8.6$M_{\oplus}$ \citep{kubfosmus19}, safely above our theoretical survival core mass.\footnote{It is also possible that at these large periods ($\gtrsim$30 days), upper atmospheres may not remain in the fluid limit out to the sonic point, and so mass loss occurs via Jeans escape. Detailed radiation-hydrodynamic calculations are needed to assess this possibility.} The overall agreement with our model (25 of 31 planets being fully consistent at the 1$\sigma$ level, 27/31 at 2$\sigma$, and 30/31 at 3$\sigma$) suggests that our calculations correctly predict the core mass requirement of sub-Saturns to be stable against envelope mass loss.

\begin{figure*}[!htb]
\centering
\includegraphics[width=1\textwidth]{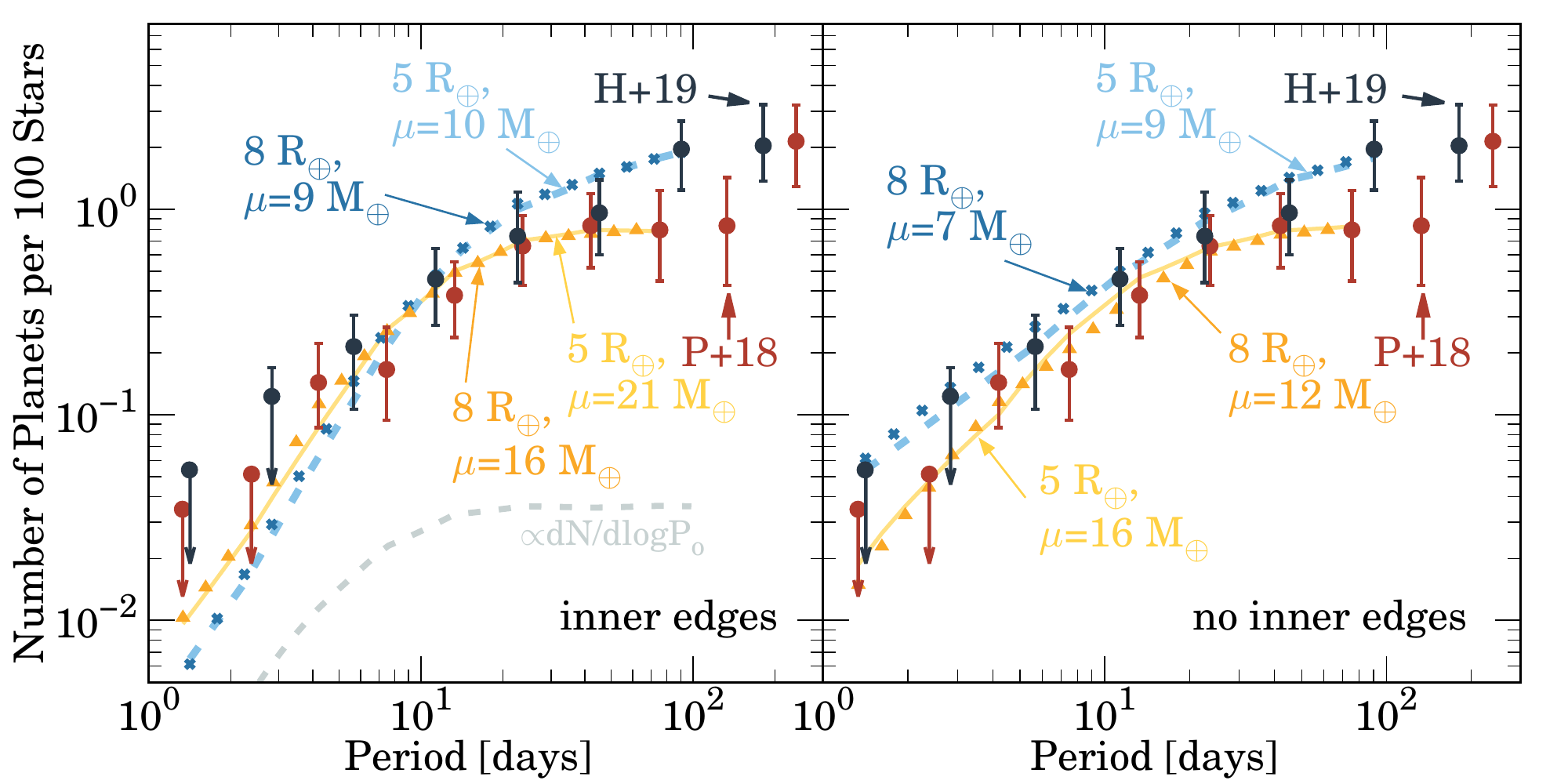}
\caption{Model occurrence rates of sub-Saturn exoplanets. Data points from \cite{petmarwin18} are in red, and data from \cite{hsuforrag19} are in dark navy. ``8 $R_{\oplus}$" and ``5 $R_{\oplus}$" denote populations representative of large and small sub-Saturns (that begin with initial gas to core mass ratios $\sim$1 and $\sim$0.2, respectively), which bracket the sub-Saturn parameter space. The left panel accounts for a decrease in occurrence inside $\sim$10 days due to truncation of disk inner edges (the initial occurrence profile is shown in the grey dashed line, normalized in this plot so as not to interfere with the model curves), while the right panel does not. These models assume lognormal core mass functions with median $\mu$ (equation \ref{equation:lognormal}; see the text for how the standard deviation $\sigma_{\rm M}$ varies between models). Model curves are normalized to the data inside 100 days, where mass loss likely operates. Mass loss is computed using the empirical model outlined in Section \ref{sec:empirical}. The sub-Saturn period distribution inside $\sim$100 days can be sculpted by atmospheric mass loss; this implies a core mass function peaked near $\sim$10--20$M_{\oplus}$.}
\label{figure:result}
\end{figure*}

Orbital period distributions based on the best-fit underlying core mass distributions of sub-Saturns are showcased in Figure \ref{figure:result} in cases where disk inner edges are accounted for (left panel) or not (right panel).

Results for the ensemble of large sub-Saturns are indicated by ``$8R_{\oplus}$", and results for the small sub-Saturns by ``$5R_{\oplus}$". Since our ensembles of large and small sub-Saturns bracket the entire sub-Saturn population, in reality the core mass function is likely a middle ground between their respective results. We control for observational uncertainties associated with the occurrence rate (as it is presently measured) by fitting \cite{hsuforrag19} and \cite{petmarwin18} separately. Fitting \cite{kunmat20} requires only slightly more bottom-heavy core mass distributions than \cite{hsuforrag19} (by $\sim$1$M_{\oplus}$), with the caveat that our model curves run outside the error bars for the data point at $\sim$4 days. This is due to the modest uptick in occurrence they report from $\sim$9--4 days, a feature which is inconsistent with our mass loss model (which always predicts decreasing occurrence with decreasing period). Since the results of this investigation serve as a proof of concept---rather than a final, rigorous determination of the core mass function---we 
only plot our fits to \dblcit{petmarwin18}{hsuforrag19} in Figure \ref{figure:result}, with the understanding that \cite{kunmat20} yield similar core mass functions to \cite{hsuforrag19}, and therefore marginalizing across all three data sets will not result in a mass function significantly different than those shown in Figure \ref{figure:result}.

We normalize our model to the data at periods $\lesssim$100 days. Normalizing to the outermost data points at 181 days for \cite{hsuforrag19} and 237 days for \cite{petmarwin18} is not physically well-motivated. Inside $\sim$100 days, hydrodynamic mass loss may operate and carve out the decrease in occurrence reported by \cite{petmarwin18,hsuforrag19,kunmat20}. Features in the occurrence rate beyond $\gtrsim$100 days however are likely sculpted by fundamental planet formation processes, rather than atmospheric mass loss (which likely ceases to be hydrodynamic at these large distances, giving way to less efficient Jeans escape). \cite{chaleeknu21} recently showed that massive planet growth through gas accretion onto cores is particularly favorable between $\sim$1 and 10 au in protoplanetary disks, due to a significant decrease in dust-to-gas ratio, and therefore opacity, just beyond the ice line that shortens envelope cooling timescales. This is in line with an observed global peak in giant planet occurrence at these distances \citep[e.g.][]{cumbutmar08,fermulpas19,barlafart19,niedermac19,fulroshir21}, and points strongly toward the physics of gas accretion playing a key role in setting the occurrence there (rather than post-processing through atmospheric mass loss).

\begin{deluxetable*}{cCCCCC}
\tablecaption{``Best-Fit" Core Mass Functions. \label{tab:results}}
\tablecolumns{5}
\tablewidth{0pt}
\tablehead{
\colhead{Mass Loss Model} &
\colhead{Occurrence Rate Data} &
\colhead{Size of sub-Saturns ($R_{\oplus}$)} &
\colhead{$\mu$ ($M_{\oplus}$)} & 
\colhead{$\sigma_{\rm M}$} &
}
\startdata
HBA & \text{\cite{petmarwin18}} & 8 & 16 & 0.7 \\
HBA & \text{\cite{petmarwin18}} & 5 & 21 & 0.7 \\
HBA & \text{\cite{hsuforrag19}} & 8 & 9 & 0.7 \\
HBA & \text{\cite{hsuforrag19}} & 5 & 10 & 0.8 \\
EL$+$P & \text{\cite{petmarwin18}} & 8 & 10 & 0.7 \\
EL$+$P & \text{\cite{petmarwin18}} & 5 & 13 & 0.7 \\
EL$+$P & \text{\cite{hsuforrag19}} & 8 & 6 & 0.7 \\
EL$+$P & \text{\cite{hsuforrag19}} & 5 & 7.5 & 0.7 \\
\enddata
\tablerefs{``HBA" and ``EL+P" refer to the empirical and analytic mass loss models outlined in Sections \ref{sec:empirical} and \ref{sec:analytic}, respectively. All referenced models include truncation of disk inner edges. Our fiducial results are based on the empirical mass loss model. ``Size of sub-Saturns" refers to the 5 Gyr radii of our ensembles of planets, in the absence of mass loss (a proxy for their initial GCR; see Section \ref{subsec:estimation_method}).}
\end{deluxetable*}

Our ``best-fit" core mass functions are summarized in Table \ref{tab:results}. Fitting \cite{petmarwin18} while accounting for the truncation of disk inner edges requires a mean $\mu=21M_{\oplus}$ with standard deviation $\sigma_{\rm M}=0.7$ for the small sub-Saturns, or $\mu=16M_{\oplus}$ and $\sigma_{\rm M}=0.7$ for the large. Neglecting the truncation of disk inner edges requires more narrow and bottom-heavy distributions, since the decrease in occurrence inside $\sim$10 days must then be due solely to mass loss. In this case, $\mu=12M_{\oplus}$ and $\sigma_{\rm M}=0.5$ for large sub-Saturns, while $\mu=16M_{\oplus}$ and $\sigma_{\rm M}=0.5$ for the small. Re-normalizing the model to match \cite{hsuforrag19} at $\sim$100 days and accounting for disk inner edge truncation requires $\mu=10M_{\oplus}$ with $\sigma_{\rm M}=0.8$ for small sub-Saturns, and $\mu=9M_{\oplus}$ with $\sigma_{\rm M}=0.7$ for the large population. The drop in occurrence from $\sim$100--50 days reported by \cite{hsuforrag19} and \cite{kunmat20} compared to the plateau at these distances reported by \cite{petmarwin18} is responsible for the difference in ``best-fit" core mass functions. Normalizing our fit to \cite{petmarwin18} at 234 days results in broader and more bottom-heavy mass functions, with $\mu=7M_{\oplus}$ and $\sigma_{\rm M}=1$ (for small and large sub-Saturns, and accounting for disk edges). While this is closer in agreement with the results from \cite{hsuforrag18}, our fiducial results should be more reliable, since it is unlikely that atmospheric mass loss is responsible for the increase in occurrence beyond $\sim$100 days. The difference between occurrence rate measurements therefore has the largest impact on our results. Future occurrence studies characterizing periods between $\sim$50--100 days would be helpful to constrain our predictions.

Our analytic model results in more bottom-heavy mass functions. Fitting \cite{hsuforrag19} requires $\mu=6-7.5M_{\oplus}$ and $\sigma_{\rm M}=0.7$, as opposed to our fiducial $\mu=9-10M_{\oplus}$ and $\sigma_{\rm M}=0.8$. \cite{petmarwin18} is recovered with $\mu=10-13M_{\oplus}$ and $\sigma_{\rm M}=0.7$, lower than our fiducial $\mu=16-21M_{\oplus}$ and $\sigma_{\rm M}=0.7$. These $\pm$3--8$M_{\oplus}$ error bars in median core mass are subsumed by the uncertainties in the occurrence rate, which can change our results by as much as 11$M_{\oplus}$ (at maximum). Our conclusions are therefore only weakly sensitive to the choice of mass loss model. Finally we note that the energy-limited photoevaporation model is difficult to reconcile with the drop in occurrence between $\sim$50--100 days reported by \cite{hsuforrag19} and \cite{kunmat20}, since pure photoevaporation results in modest mass loss rates at these large periods. 

\begin{figure}
\epsscale{1.2}
\plotone{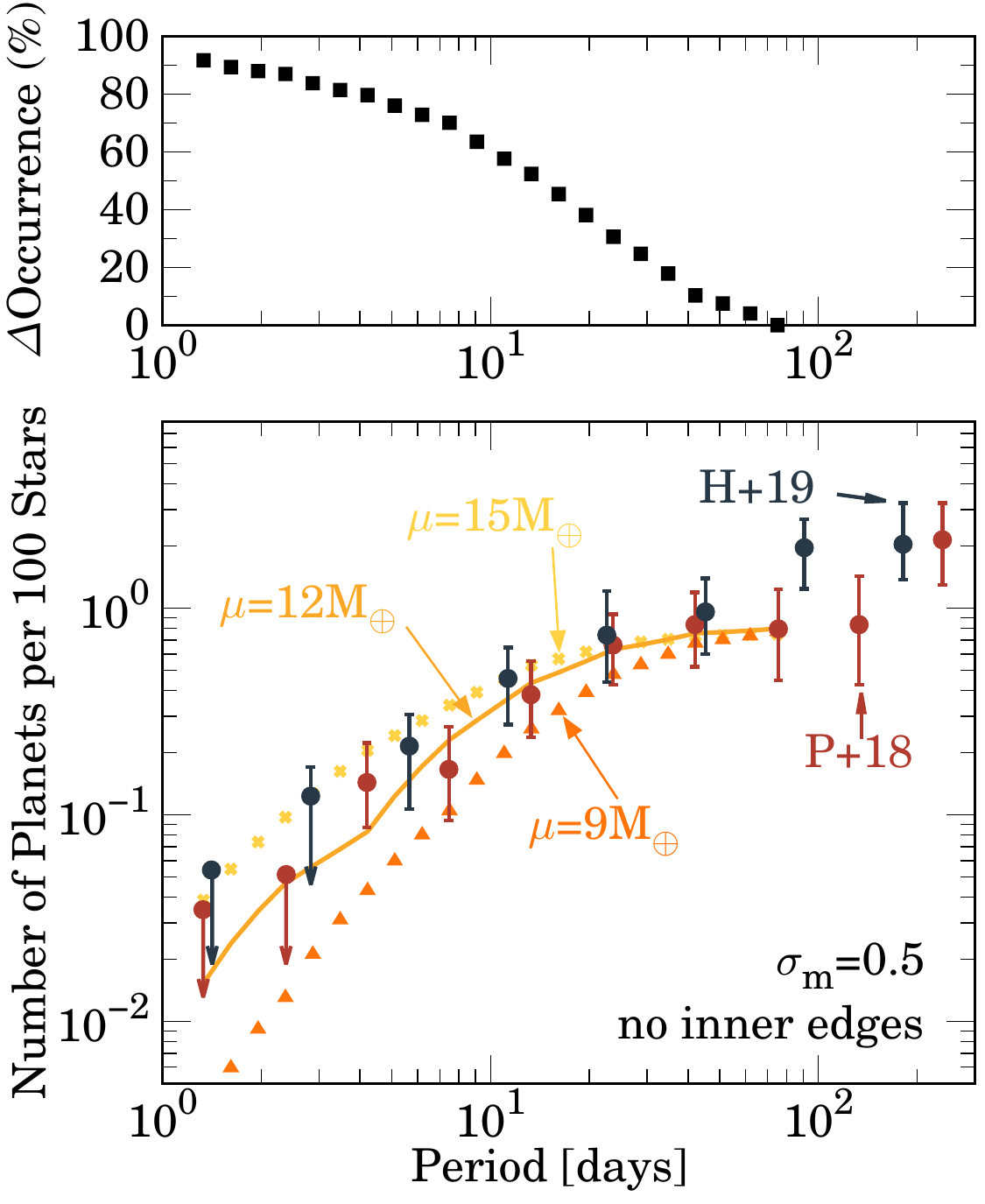}
\caption{Top: percentage decrease in occurrence as a function of orbital period between core mass functions with medians $\mu$=15$M_{\oplus}$ versus $\mu$=9$M_{\oplus}$ (both with $\sigma_{\rm M}$=0.5, and $P_{\star}$=10 days). The occurrence decreases with lower $\mu$ because atmospheres can be stripped more readily around lower mass cores. Bottom: model occurrence rates fit to the occurrence data from \cite{petmarwin18} using different core mass functions. To isolate the effect of the core mass function, these models do not account for disk inner edge truncation. Occurrence rates are shown using median core masses $\mu\pm$1.8$\sigma_{\rm M}$ above and below the ``best-fit" value $\mu$=12$M_{\oplus}$ ($\sigma_{\rm M}$=0.5 or $\sim$1.65$M_{\oplus}$, dark orange, solid curve), with $\mu$=15$M_{\oplus}$ (marked by light orange crosses), and 9$M_{\oplus}$ (darkest orange, triangles). Only the ``best-fit" mass function with $\mu$=12$M_{\oplus}$ can recover the observed occurrence rate within its 1$\sigma$ error bars. The sub-Saturn orbital period distribution can constrain the sub-Saturn core mass distribution to within $\sim$1.8$\sigma_{\rm M}$. \label{fig:evap_mcore}}
\end{figure}

Figure \ref{fig:evap_mcore} illustrates how the occurrence rate evolves with the choice of the underlying core mass function. We plot an example ``best-fit" core mass function with $\mu$=12$M_{\oplus}$, along with mass functions $\pm$3$M_{\oplus}$ larger or smaller in $\mu$. Only the ``best-fit" model successfully recovers the data within the 1$\sigma$ error bars, and the occurrence can decrease by as much as $\sim$90\% across the models with the largest and smallest median core masses. The occurrence rate strongly constrains the underlying core mass function.

\subsection{Super-Puffs}\label{sec:super-puffs}
Our model predicts that two Super-Puff planets should not presently be observed: Kepler-33 d ($3.91^{+1.87}_{-1.8}M_{\oplus}$, $4.58^{+0.96}_{-0.96}R_{\oplus}$; \citealt{hadlit16}) and Kepler-223 d ($8.0^{+1.5}_{-1.3}M_{\oplus}$, $5.24^{+0.26}_{-0.45}R_{\oplus}$; \citealt{millfabmig16}). Their low densities indicate that their atmospheres are especially susceptible to rapid mass loss, and yet they have persisted for Gyr to the present. Despite the fact that our comparison with \cite{milpetbat20} accounts for their larger core mass estimates due to tidal heating, these planets still exhibit perplexingly short atmospheric lifetimes. 
Proposed solutions suggest that Super-Puffs have overestimated transit radii and therefore artificially low densities, either due to high-altitude aerosols (\cite{lamerkfos16,cuberkjuv17}; see also \cite{wandai19,gaozha20,ohntan21} for dust advection and cloud formation models), or planetary ring systems \citep{pirvis20}. Here we test an alternative explanation: they began with even more massive envelopes, and are presently in the process of losing them. Despite their small masses, these cores are capable of accreting large amounts of gas provided they begin at large ($\sim$au) distances where envelopes can cool more quickly,
either in environments where dust grains do not contribute to the opacity \citep{leechi16} or in environments where dust grains drift in to the inner disk pile-up creating a localized region of low dust-to-gas ratio and therefore opacity \citep{chaleeknu21}. 
While this hypothesis does require that gas accretion is halted in the midst of runaway, 
we show below that it is not ruled out by the observations.

Super-Puffs are especially susceptible to mass loss in the form of a Parker wind. As pointed out by \cite{owewu16}, in the limit that $R_{\rm p} \ll R_{\rm B}$, the Mach number $\sqrt{-W_{0}(-f(R_{\rm p}/R_{\rm B}))}\sim (R_{\rm B}/R_{\rm p})^{2}\exp{(-2R_{\rm B}/R_{\rm p})}$ (since to first order $W_{0}(x)\sim x$ for $|x|\ll$1 \citep[e.g.][]{corgonhar96}), meaning the mass loss rate is exponentially sensitive to the planetary and Bondi radii. Super-Puffs' large radii and low masses make the fraction $R_{\rm p}/R_{\rm B}\gtrsim$0.1 non-negligible, especially at early times in our simulations when their radii are inflated. Rapid destruction of their atmospheres can therefore be avoided if the planet mass can be increased (thereby increasing $R_{\rm B}$) without increasing the planetary radius. Above GCRs $\sim$1, envelopes' self-gravity leads to only modest increases in radii with mass (at fixed core mass). We tested this hypothesis by constructing a sample of planets with GCRs spanning $\Gamma=0.2-4.0$ surrounding a $10M_{\oplus}$ core at 0.1 au. We observed (after 10 Myr of thermal evolution) an approximate power-law relation $R_{\rm p}\propto M^{\sim 1.4}_{\rm p}$ for $\Gamma\lesssim$1, and a flattening of the radius-mass relation above $\Gamma\sim1$ where $R_{\rm p}\propto M^{\sim 0.4}_{\rm p}$ for $1\lesssim \Gamma\lesssim$2 and $R_{\rm p}\propto M^{\sim 0.1}_{\rm p}$ for $\Gamma\gtrsim$2. The empirical relation derived by \cite{chekip17} gives $R_{\rm p}\propto M^{0.59}_{\rm p}$ for Uranus and Neptune-like planets ($\sim$4$R_{\oplus}$, $\sim$15$M_{\oplus}$), a middle ground between our results for small and large envelopes (this difference for $\Gamma\lesssim$1 is expected, given that we compute our mass-radius relation at 10 Myr instead of 5 Gyr). Super-Puffs with initial GCRs $\gtrsim$1 reach $\sim$10$R_{\oplus}$ (after $\sim$Gyr of thermal contraction), placing them at the upper limit of planetary radii, even for those near the transition regime between Neptunes and Jovians \citep{chekip17}. Since, for these large envelopes, the Bondi radius scales more strongly with planet mass ($R_{\rm B}\propto M_{\rm p}$) than the planetary radius does, initially large GCRs $\Gamma\gtrsim$1 can push the Bondi radius far enough out that the Parker wind effectively shuts off ($R/R_{\rm B}<0.1$).

\begin{figure}
\epsscale{1.2}
\plotone{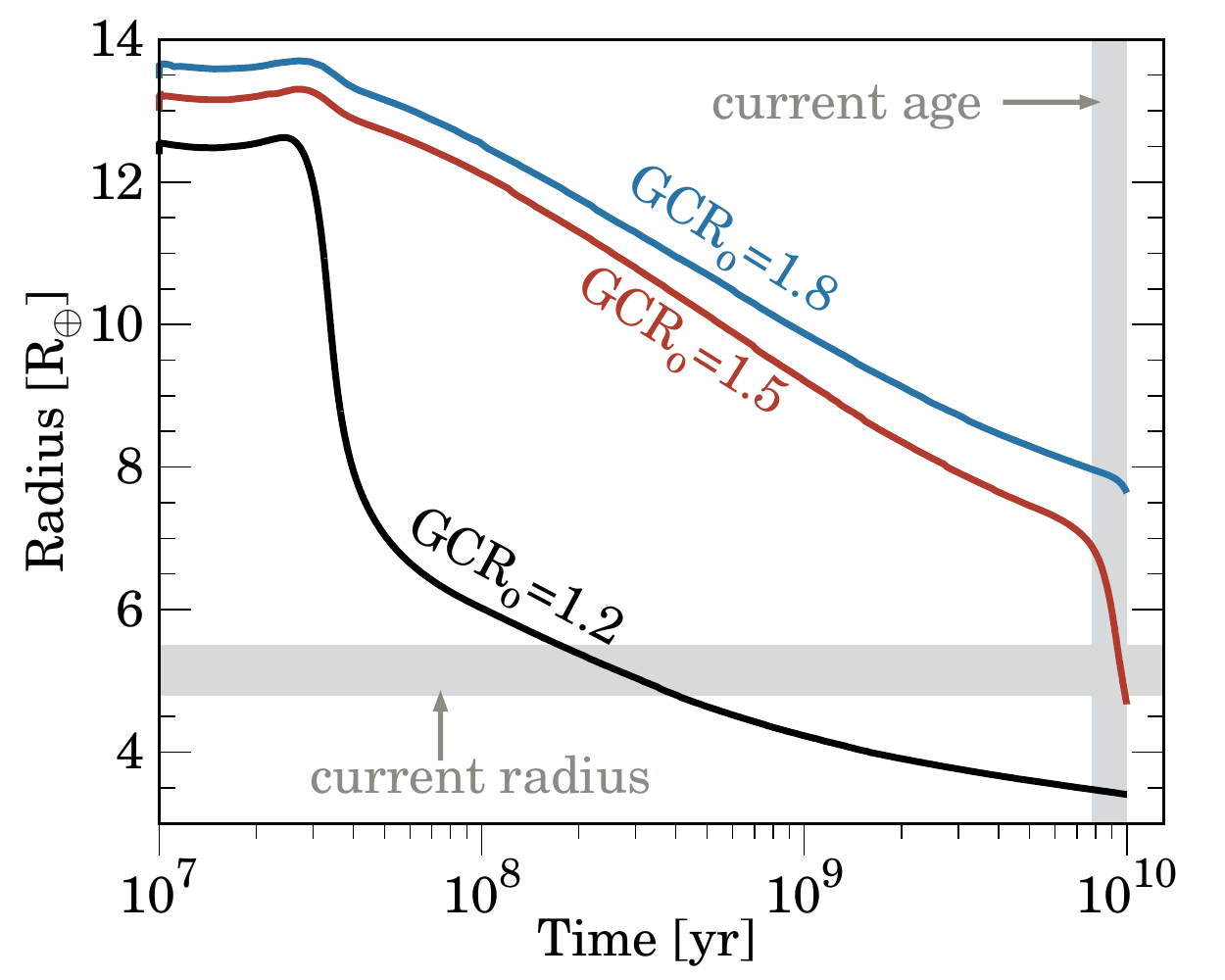}
\caption{Possible radius evolution of Kepler 223-d, for different values of the initial gas to core mass ratio ``GCR$_{o}$" ($=1.2$ in black, $=1.5$ in red, and $=1.8$ in blue). We adopt a core mass of $7.7M_{\oplus}$ (consistent with the upper limit without tides from \cite{milpetbat20}), a stellar mass and age of $1M_{\odot}$ and $\sim$9 Gyr (from \cite{fulpet18}; the present stellar age is shaded in the vertical grey line), and the present-day radius of Kepler 223-d from \cite{millfabmig16} (shaded in the horizontal grey line). The uptick in radius at $\sim$30 Myr is due to a local maximum in stellar bolometric luminosity at the end of the pre-main sequence, which inflates the upper atmosphere. The smallest initial envelope in black triggers rapid hydrodynamic escape through a Parker wind outflow that shrinks the radius well below that observed; the largest initial envelope in blue pushes the Bondi radius so far beyond the planet radius that the Parker wind is effectively shut off, and Kepler 223-d instead passively photoevaporates; and the fiducial envelope in red undergoes late-time thermal escape catalyzed by an increase in the stellar bolometric luminosity as it approaches the end of the main sequence. In this picture, Kepler 223-d is currently losing $\sim$2$\times 10^{-3}M_{\oplus}\mathrm{Myr}^{-1}$.
Initial envelopes even larger than in the present can help Super-Puffs avoid rapid mass loss. \label{fig:superpuff}}
\end{figure}

In this picture, the ultimate fate of a Super-Puff is sensitive to its initial GCR. This is illustrated in Figure \ref{fig:superpuff}, where we display a case study using Kepler 223 d. Because its initial radius is well outside the hydrodynamic approximation boundary, we compute the mass loss history with our analytic model. Choosing $\Gamma$ too small (e.g. $1.2$, black curve in Figure \ref{fig:superpuff}) leads to rapid hydrodynamic escape catalyzed at $\sim$20 Myr by a local maximum in stellar bolometric luminosity at the end of the pre-main sequence, and too large an envelope (e.g. $\Gamma=1.8$, blue curve) can allow the planet to survive without undergoing any degree of thermal escape (instead passively photoevaporating for Gyr timescales).

As shown in Figure \ref{fig:superpuff}, we find that initial GCRs $\sim$1.5 successfully reproduce Kepler 223-d's measured radius and mass in the present day (using values for the radius and mass from \cite{millfabmig16}, stellar age and mass from \cite{fulpet18}, and core mass estimates from \cite{milpetbat20}). Ordinary photoevaporation dominates for the bulk of its evolution, while the stellar bolometric luminosity is essentially constant. By $\sim$9 Gyr (the estimated age of the system \citep[][]{fulpet18}), the host star is nearing the end of the main sequence, and its increasing bolometric luminosity heats up the planet's photosphere. At this point the Bondi radius has shrunk from its initial value by a factor of $\sim$2, partly due to the increase in photospheric sound speed, and partly due to the steady decrease in mass the planet experiences via photoevaporation. This fuels a late stage of enhanced mass loss when $R/R_{\rm B}\sim$0.09 reaches a local maximum. In this picture, Kepler 223 d is currently losing $\sim$4$\times 10^{11}\mathrm{g~s^{-1}}$ or $2\times 10^{-3}M_{\oplus}\mathrm{Myr^{-1}}$. A similar hypothesis regarding a late-stage of mass loss for 223 d was also recently mentioned by \cite{chajonknu20}. We defer a more careful treatment of the problem (e.g., MCMC sampling of the stellar and planet initial conditions) to future work. 

Kepler 223 e's low density (4.8$^{+1.4}_{-1.2}M_{\oplus}$, 4.6$^{+0.27}_{-0.41}R_{\oplus}$ \citep{millfabmig16}) is likely also consistent with late-time boil-off. Future studies of the Kepler 223 resonant chain's long-term stability should account for late-time mass loss processes, which can alter the dynamics of resonant systems \citep[][]{matogi20}.

A similar analysis for Kepler 33 d reveals that in this picture it should presently be undergoing strong hydrodynamic escape, currently losing $\sim$5$\times 10^{13}\mathrm{g~s^{-1}}$. Given this high mass loss rate, we find that it should only be observable at its present radius and mass for a short window lasting $\sim$10 Myr. High-altitude aerosols (or circumplanetary rings) may provide a more robust explanation for Kepler 33 d.

\subsection{Sub-Neptunes That Begin as sub-Saturns}

Defining sub-Neptunes as planets with radii in the range 1--4$R_{\oplus}$, \footnote{Planets $\lesssim$4$R_{\oplus}$ fall into the Super-Earth ($\sim$1--1.7$R_{\oplus}$) or sub-Neptune ($\sim$1.7--4$R_{\oplus}$) categories, but 
this distinction is irrelevant for our purposes as 
we are interested in the survival of sub-Saturns.} the observed ratio of the number of sub-Neptunes to sub-Saturns ranges from $\sim$10--50, depending on orbital period and the occurrence measurement \citep[][]{petmarwin18,hsuforrag19,kunmat20}.
The ratio between sub-Saturns that transform into sub-Neptunes and sub-Saturns that survive mass loss predicted from our model accounts for $\sim$10--20\% of the measured sub-Neptune/sub-Saturn fraction at $\sim$4 days, and falls to $\sim$1--5\% from $\sim$10--30 days. Our model predicts that $\sim$5--10\% of the total sub-Neptunes between 4 and 20 days were once sub-Saturns that have since lost an appreciable fraction of their atmospheres. 

\begin{figure}
\epsscale{1.2}
\plotone{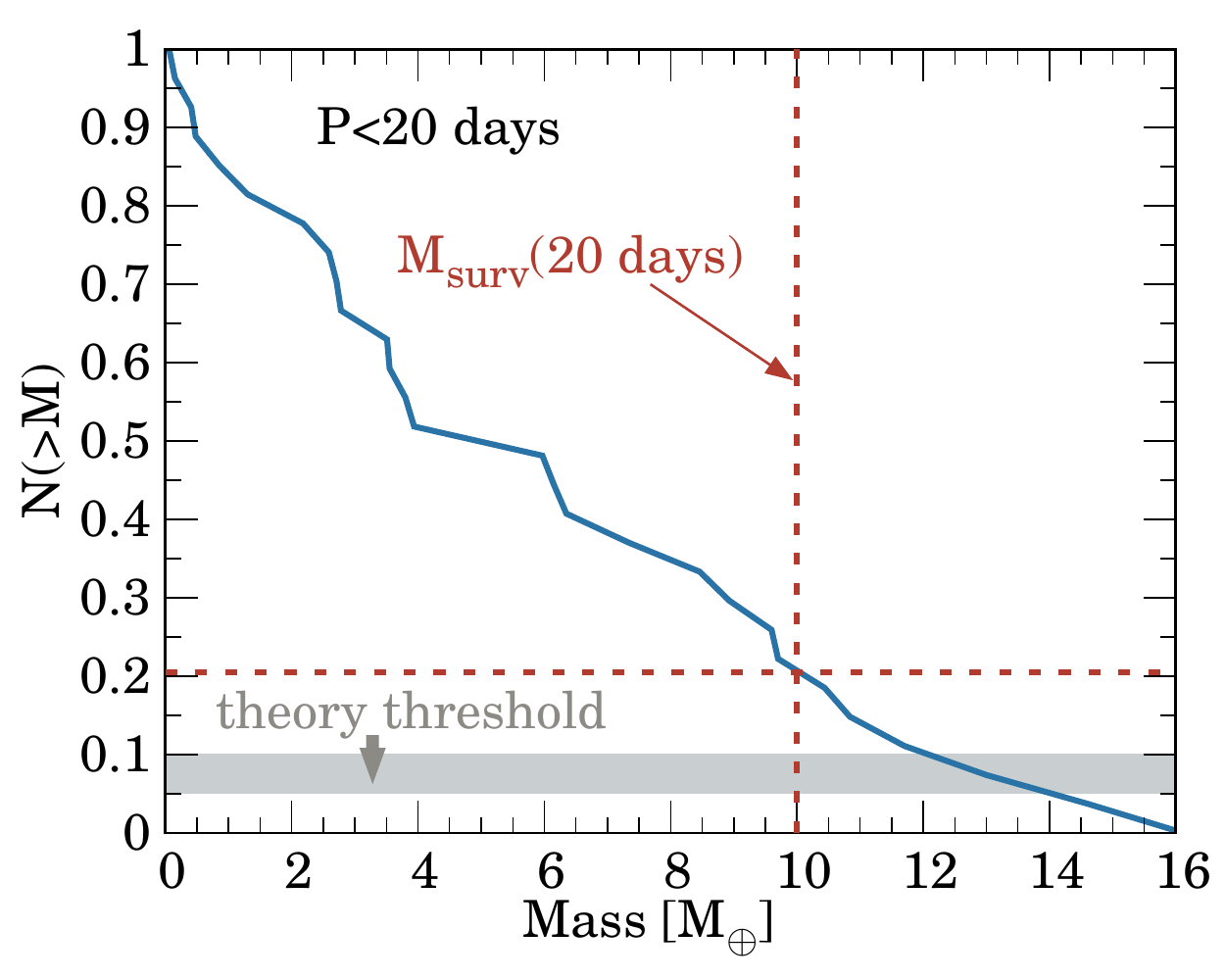}
\caption{The mass distribution of close-in sub-Neptunes as measured by the radial velocity follow-up study of \cite{marisahow14}. We have kept only those planets from \cite{marisahow14} that have mass measurement uncertainties, positive mass estimates, radii $<$4$R_{\oplus}$, and periods $<$20 days. The grey horizontal shaded area demarcates the $\sim$5--10\% threshold of warm sub-Neptunes that our mass loss model predicts began as more massive sub-Saturns. The red dashed lines indicate the $\sim$21\% of sub-Neptunes in the sample that have masses consistent with a possible sub-Saturn origin (i.e. have masses smaller than the survival core mass at their respective periods, but larger than the survival core mass at 20 days, since inside 20 days cores even larger than this can still lose significant atmosphere and become sub-Neptunes). This is consistent with our model prediction. \label{fig:sncdf}}
\end{figure}

This prediction is corroborated by the radial velocity follow-up measurements of Kepler sub-Neptunes from \cite{marisahow14}. The minimum core mass at which we expect warm sub-Neptunes (inside orbital periods $\sim$20 days) to begin as sub-Saturns is approximately the survival core mass at 20 days (for our ``small" sub-Saturns, as most sub-Neptunes that began as sub-Saturns are more likely to have formed with GCRs $\sim$20\% rather than $\sim$100\%), since inside 20 days cores even larger than this threshold could still lose significant amounts of envelope to become sub-Neptunes. Taking the total measured mass of sub-Neptunes to be that of their cores, sub-Neptunes that may have originated as sub-Saturns must be lighter than the survival core mass at their respective periods. We note that it is also possible that sub-Neptunes that fit this mass criteria may have been born as massive sub-Neptunes. In Figure \ref{fig:sncdf}, we compare the model's expected fraction of sub-Neptunes with a sub-Saturn origin ($\sim$5--10\%, highlighted in grey) with the fraction of observed sub-Neptunes with masses consistent with a possible sub-Saturn origin (6 planets of 28 total in the sample, or $\sim$21\%, highlighted with red dashed lines). All 6 of these planets (specifically: Kepler 94 b, Kepler 113 b, Kepler 48 c, Kepler 95 b, Kepler 106 c, and Kepler 131 b) are indeed smaller than the survival core mass at their respective periods, 
within their 1$\sigma$ error bar.
%This corroborates our model prediction of $\sim$5--10\%, and we are assured that it is not an over-prediction. 
The observed number fraction of sub-Neptunes with potential sub-Saturn origin $\sim$21\% is higher than the model prediction $\sim$5--10\% which is consistent with our expectation that the observed population includes sub-Neptunes that may have born as massive sub-Neptunes.
A caveat to this comparison however is that we have not corrected for observational biases against low mass objects.

\section{Discussion}\label{sec:discussion}

\subsection{Observational Predictions}

\begin{figure}
\epsscale{1.2}
\plotone{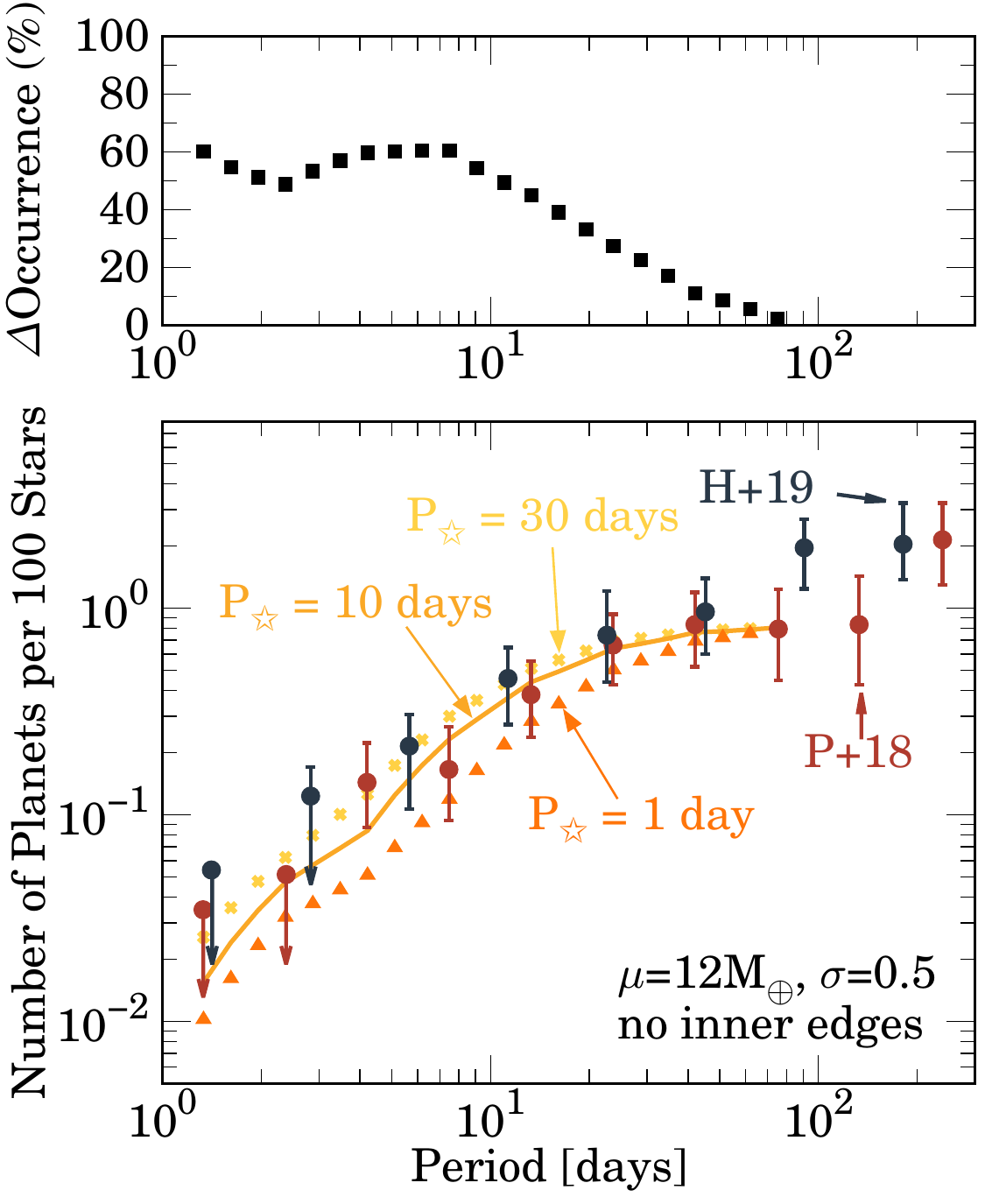}
\caption{Top: similar to Figure \ref{fig:evap_mcore}, the percentage decrease in occurrence as a function of orbital period but between models using an initial stellar rotation period $P_{\star}$=30 days versus those using $P_{\star}$=1 day. The occurrence decreases with shorter initial stellar rotation periods due to the increase in XUV output during their prolonged saturation phase. Bottom: model occurrence rates fit to the occurrence data from \cite{petmarwin18} using different initial stellar rotation periods. To isolate the effect of stellar rotation, these models do not account for disk inner edge truncation. Occurrence rates are shown using using $P_{\star}$=30 days (marked by light orange crosses), 10 days (darker orange, solid curve), and 1 day (darkest orange, triangles). Close-in Sub-Saturns should be $\sim$50\% less common and $\sim$30\% more massive around rapidly rotating stars. \label{fig:evap_pstar}}
\end{figure}

Slowly rotating stars can fall out of XUV saturation as much as an order of magnitude earlier than fast rotators \citep[e.g.][]{tujohgud15,johbargud20}. We therefore predict that the number of close-in (i.e. periods $\lesssim$10 days) sub-Saturns should increase with increasing stellar rotation period, for stars of similar metallicity and mass.\footnote{\cite{amamat20} highlight a non-linear dependence of the rotation evolution---and by extension the XUV luminosity evolution---on the metallicity, such that low-metallicity stars should transition out of saturation earlier than high-metallicity stars with the same initial conditions.} In Figure \ref{fig:evap_pstar} we plot an example occurrence rate model fit to \cite{petmarwin18} using different initial stellar rotation periods; we observe a $\sim$30--60\% decrease in occurrence from slow to fast rotators inside periods of $\sim$20 days. Evidence of this effect may already be present in the Kepler data; \cite{mcqmazaig13} report that only stars with rotation periods longer than $\sim$5--10 days host planets inside $\sim$3 days, and the enhanced mass loss around rapid rotators may at least partially contribute to this. Similarly, close-in sub-Saturns found around fast rotators (i.e. those that have survived mass loss) should be more massive than those orbiting slow rotators (at fixed stellar metallicity and mass). We find that $M_{\rm surv}$ increases by $\sim$15--30\% when decreasing $P_{\star}$ from 10--1 days, indicating that sub-Saturns orbiting fast rotators should be $\sim$15--30\% more massive than those found around slow rotators. This effect will obviously be easier to discern for younger stars that have not spun down significantly. Observations of open clusters as well as interior rotation models both indicate that young stars' rotation periods converge by $\sim$500 Myr \citep[e.g.][]{irwbou09,galbou13}, so it should therefore still be possible to discern between these populations for systems younger than $\sim$500 Myr.

Photoevaporation has been proposed as the physical origin of the lower boundary of the sub-Jovian desert \citep[][]{owelai18}. In this picture, photoevaporation sets an upper limit on the radii and masses of sub-Neptunes
at very small orbital periods, such that planets with radii and masses between $4R_{\oplus}\lesssim R_{\rm p}\lesssim$10$R_{\oplus}$ and $10M_{\oplus}\lesssim M_{\rm p}\lesssim$100$M_{\oplus}$ are rarely found inside periods of $\sim$3 days. We make the ansatz that the underlying core mass function of sub-Saturns---a lognormal peaked near 10--20$M_{\oplus}$---is universal, applying across variations in stellar mass and orbital period. We can estimate empirically from Figure \ref{figure:subsaturn_occurrence} that the desert begins where the occurrence rate $\sim$0.12 (at periods $\sim$3 days). Figure \ref{figure:mcritical} reveals that the corresponding survival core mass (which yields a $10\%$ survival probability) is $\sim$22$M_\oplus$ and $\sim$32$M_{\oplus}$, for large and small sub-Saturns respectively. The desert should therefore begin at shorter orbital periods around lower mass stars: $\sim$1.5 days around 0.8$M_\odot$ stars and $\sim$0.7 days around 0.5$M_\odot$ stars. The lower boundary of the desert should also decrease towards lower maximum masses and radii with increasing stellar mass: the desert should grow wider due to the increase in XUV and bolometric emission.

This shift in the opening and width of the sub-Jovian desert should therefore be most easily discernible when comparing between samples of M dwarfs and $\sim$1$M_{\odot}$ stars (this assumes that the sub-Saturn formation efficiency does not vary with stellar mass, which may not be true). In general, initial rotation rates must be taken into consideration in searching for trends of sub-Jovian desert morphology with host stellar type as any differences in XUV emission intrinsic to stellar mass can be erased by differences in initial rotation period. The XUV evolution between M dwarfs ($\lesssim$0.5$M_{\odot}$) and solar-type stars are sufficiently distinct however, just from stellar mass alone, that the differences in initial rotation matter little.

There are encouraging empirical results that corroborate our prediction of a wider sub-Jovian desert around more massive stars. \cite{szakal19} report a significant increase in the occurrence of close-in planets with decreasing stellar effective temperature (for effective temperatures $<$5600 K). They find that $\sim$60\% of planets orbiting stars with effective temperatures $<$5600 K are found inside 10 days, as opposed to the $\sim$10\% of planets around hotter stars found inside 10 days. Since effective temperature scales with stellar mass (and stars less massive than $1M_{\odot}$ have temperatures $\lesssim$5600 K), we interpret this as a possible indication that the desert around lower mass stars is smaller in extent and may begin at smaller periods. Because sub-Neptunes are more common around cooler stars \citep[e.g.][]{drecha15}, determining whether this feature is a result of mass loss and not intrinsic planet occurrence will require more detailed analysis of the mass/radius vs. period distribution as a function of stellar type to tease out possible signatures of the desert and its boundaries.

\subsection{Limitations of the Model $\&$ Future Work}\label{sec:caveats}

Our structure models necessarily simplified some of the physical processes at work in the interiors of massive planets. We outline below the most pertinent simplifications our models have made, and to what extent a more detailed treatment would affect our results.

Hot and warm sub-Saturns will inevitably experience strong tidal effects from interaction with their host stars. Often found with moderate eccentricities \citep[][]{petsinlop17}, they likely experience eccentricity and obliquity tides \cite[e.g.][]{milpetbat20}. Tidal dissipation could affect our model in two ways: by altering the mass loss process, and by driving orbital migration due to tidal inspiral. \cite{milpetbat20} found that tidal radius inflation may imply $\sim$10\% lower atmospheric mass fractions than in conventional structure models that neglect tides (such as ours; in some cases, envelope fractions can deviate up to $\sim$25\%). It is unclear whether adding tidal dissipation may lead to more or less atmospheric mass loss; tidal heating implies more massive cores, which stabilize against atmospheric mass loss, but conversely the extra internal luminosity (if strong enough) could enhance boil-off at early times. More detailed models self-consistently coupling tidal heating and atmospheric mass loss (also accounting for changes in orbital distance due to non-zero eccentricities) would be helpful. Orbital inspiral could decrease our model occurrence rates inward of $\sim$5 days; using the tidal dissipation formalism of \cite{han10} with $Q_{\rm p}=10^{6}$ ($\bar{\sigma}_{\rm p}=3.4\times 10^{-5}$) and $Q_{\star}=10^{6}$, we find that typical sub-Saturns inside $\sim$5 days (i.e. which need to be $\gtrsim$20$M_{\oplus}$ to remain sub-Saturns after mass loss, and will have typical radii $\sim$8$R_{\oplus}$) spiral inwards past the stellar Roche surface over Gyr timescales (assuming no planetary spin and zero eccentricity). Our model occurrence rates could therefore decrease inside $\sim$5 days if $Q_{\rm p}\lesssim$10$^{6}$ (larger $Q_{\rm p}\sim$10$^{7-8}$ produce negligible orbital migration).

The mass loss process can be weakened by planetary dipole magnetic fields. These can redirect outflowing winds and decrease photoevaporative mass loss by as much as an order of magnitude \citep[e.g.][]{oweada14}. In this case we would expect the sub-Saturn core mass function to be more bottom-heavy than estimated in this work. The strength of exoplanetary magnetic fields and their interaction with host stars is still an area of active research \citep[e.g.][]{ada11,oweada14,khoshalam15}.

Lastly, we emphasize that the fundamental goal of this study is to isolate the effect of mass loss on the sub-Saturn occurrence rate. Our model therefore assumed an initial period distribution that is flat in $\log P$. Sub-Saturns are generally massive enough to carve out gaps in the inner disk, so that disk migration---more precisely, the halting of migration by deep gaps---provides a plausible way to accomplish this \citep[][]{hallee20}. In reality however it will be necessary to add corrections to the period distribution due to, for example, the efficiency of core formation as a function of orbital period.

\begin{figure}
\epsscale{1.2}
\plotone{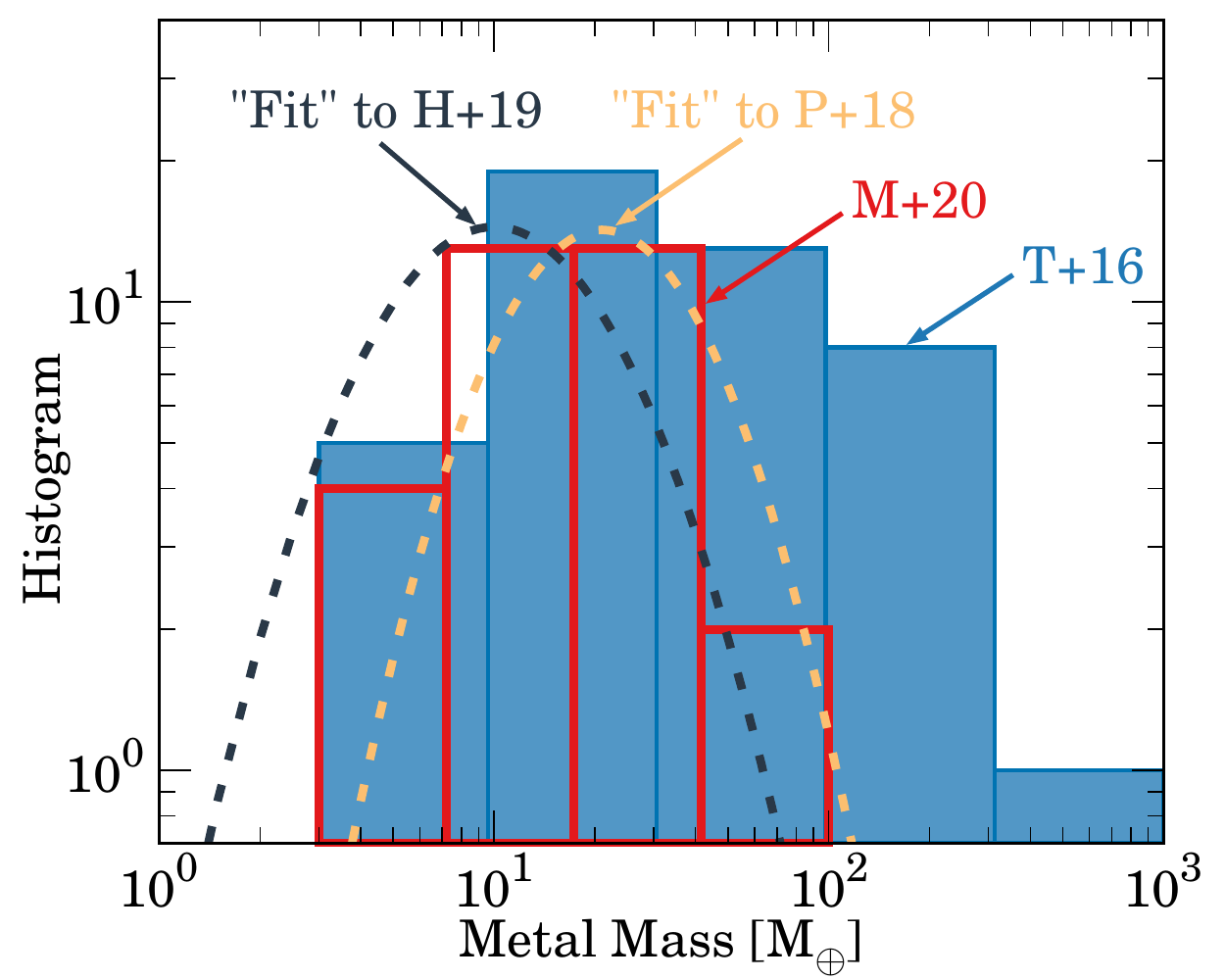}
\caption{
Metal mass distributions of gas-rich planets inferred from observations compared with our model expectations. \cite{thoformur16} (in blue) report metal masses for sub-Saturns and giant planets, accounting for up to $10M_{\oplus}$ in solid cores and the rest mixed into the envelope. \cite{milpetbat20} (in red) infer the solid core masses of sub-Saturns, accounting for tidal radius inflation. Core mass functions predicted from our mass loss theory are in dashed curves; in orange we plot the result of ``fitting" to the occurrence rate from \cite{petmarwin18} (a lognormal with $\mu=21M_{\oplus}$ and $\sigma_{\rm M}=0.7$, for our small sub-Saturns) and in dark navy we show the result of ``fitting" \cite{hsuforrag19} ($\mu=10,~\sigma_{\rm M}=0.8$). Mass functions peaked near $\sim$10--20$M_{\oplus}$ are consistent with the sub-Saturn period distribution as well as their observationally-inferred core distribution. \label{figure:corefunction}}
\end{figure}

\section{Conclusion}\label{sec:conclusion}
\subsection{Implications for Giant Planets}

The core masses of giant planets are highly uncertain. Structure models of Jupiter, for which we have the most observational constraints, predict a wide range of core masses (a spread of $\sim$20$M_{\oplus}$), depending sensitively on the equation of state as well as the degree of core dilution \cite[e.g.][]{fornet10,wahhubmil17}. Because the cores of sub-Saturns and Jupiter-sized planets likely share the same origin channel---both having reached the heavy masses required for runaway growth, sub-Saturns having prematurely halted accretion---constraints on the sub-Saturn core mass distribution may also inform that of Jupiters. 

In Figure \ref{figure:corefunction} we plot the distribution of estimated metal masses of giant planets from \cite{thoformur16} (in blue, with up to $10M_{\oplus}$ deposited in a solid core, the remaining metals mixed into the envelopes), and the estimated core masses of sub-Saturns from \cite{milpetbat20} (in red, corrected for tidal inflation). Also plotted are our theoretical core mass functions from ``fitting" \cite{petmarwin18} (in orange, a lognormal with $\mu=21M_{\oplus}$, $\sigma_{\rm M}=0.7$, for our ensemble of small sub-Saturns) and fitting \cite{hsuforrag19} (in dark navy, $\mu=10M_{\oplus}$, $\sigma_{\rm M}=0.8$). The \cite{milpetbat20} distribution exhibits a peak near $\sim$10--30$M_{\oplus}$, in reasonable agreement with our model. \cite{thoformur16} has a considerable skew towards very large metal masses, which is not surprising given that they consider giant planets with cores of maximum mass 10$M_{\oplus}$, with the rest of the metals distributed throughout the interior whether from core dilution or atmospheric pollution. A typical core mass $\sim$10--20$M_{\oplus}$ is also consistent with cores whose runaway accretion is stymied by disk hydrodynamics, such that they ultimately end with GCRs in the sub-Saturn regime $\sim$0.1--1 \citep[][see their Figure 6]{lee19}.

\subsection{Summary}

We have argued that the sub-Saturn occurrence rate may be leveraged to place novel constraints on their underlying core mass function. We summarize below our key results and their implications for gas-rich planets:

\begin{itemize}
\item The sub-Saturn period distribution inside $\sim$100 days can be sculpted by atmospheric mass loss. Photoevaporation plays the dominant role, while ``boil-off" shortly after disk dispersal may also contribute (the degree to which depends on the occurrence measurement, with \cite{hsuforrag19} requiring a larger contribution than \cite{petmarwin18}).
\item In this picture, the rise in occurrence rate with period can be used to derive the present-day sub-Saturn core mass distribution integrated over orbital periods $\lesssim$100 days. 
At shorter periods, more massive cores are required to retain their envelopes against mass loss,
so the cores of short-period sub-Saturns are sampled from the higher mass end tail of the overall distribution.
This produces the observed drop in occurrence at shorter periods.
\item While our model presented here serves as a proof of concept, we may already conclude that lognormal core mass functions peaked between $\sim$10--20$M_{\oplus}$ are compatible with the measured orbital period distribution of sub-Saturns, the distribution of observationally-inferred sub-Saturn cores, as well as gas accretion theories.
\item Some super-puffs (anomalously low density sub-Saturns) can avoid rapid destruction of their atmospheres and survive to the present day if they began with envelopes $\sim$1.5--2$\times$ the mass of their cores. In this picture, Kepler 223 d is currently undergoing a period of late-time boil-off, losing mass at a rate $\sim$2$\times 10^{-3} M_{\oplus}$Myr$^{-1}$. Late-time boil-off may alter the long-term dynamical stability of the Kepler 223 resonant chain, and other systems like it that host a super-puff.
\end{itemize}

Our mass loss theory also makes several predictions that can be tested in the near future:

\begin{itemize} 
\item The occurrence rate of close-in sub-Saturns should increase with increasing stellar rotation period (provided stellar mass and metallicity are controlled for). We estimate that sub-Saturns should be $\sim$50\% less common and $\sim$30\% more massive around rapidly rotating stars than around slow rotators. Testing this effect should be possible for stars younger than $\sim$500 Myr.
\item The sub-Jovian desert should become less pronounced around M stars. In particular, the desert should open up at smaller orbital periods around M stars compared to solar-type stars; we estimate that the desert should begin at $\sim$0.7 days around M stars, compared to opening at $\sim$3 days around solar-mass stars.
\end{itemize}

\noindent Our model predictions can be directly tested with more mass, radius, and period measurements of sub-Saturns around stars of varying types. If confirmed by the data, our predictions can also serve as a constraint on core formation theories of gas-rich planets.

\acknowledgments
T.H. thanks Daria Kubyshkina for her helpful feedback regarding the hydrodynamic approximation. We thank Gabriel-Dominique Marleau, Sarah Millholland, James Owen, and Erik Petigura for useful comments that improved the manuscript. We also thank Howard Chen for his assistance with \texttt{MESA}, and the referee for their constructive report which improved the manuscript. T.H. was supported in part by the Natural Sciences and Engineering Research Council of Canada (NSERC) through an Alexander Graham Bell Scholarship. T.H. also acknowledges support from an L. Trottier Science Accelerator Fellowship and a TEPS-CREATE Fellowship. E.J.L. gratefully acknowledges support from NSERC, le Fonds de recherche du Québec – Nature et technologies (FRQNT), McGill Space Institute, and the William Dawson Scholarship from McGill University. This research was enabled in part by support provided by Calcul Qu\'{e}bec (\url{www.calculquebec.ca}), Compute Ontario  (\url{computeontario.ca}) and Compute Canada (\url{www.computecanada.ca}).
Figures were produced using \texttt{gnuplot}.

\bibliography{hallatt}{}
\bibliographystyle{aasjournal}

\end{document}